\documentclass[12pt]{article}
\usepackage{amsmath}
\usepackage{graphicx,psfrag,epsf}
\usepackage{enumerate}
\usepackage{natbib}
\usepackage{hyperref} 
\usepackage{amsfonts}
\usepackage{amsthm}
\newtheorem{mythm}{Theorem}
\newtheorem{mydef}{Definition}

\newtheorem{myassump}{Assumption}
\usepackage{epsfig}
\usepackage{paralist}
\usepackage{fullpage}
\usepackage{setspace}
\usepackage{graphicx}
\usepackage{caption}
\usepackage{subcaption}
\usepackage{enumitem}
\allowdisplaybreaks
\usepackage{xr}
\begin{document}

\title{Robust Hypothesis Testing via L$q$-Likelihood}
\author{Yichen~Qin~and~Carey~E.~Priebe\thanks{Yichen Qin is Assistant Professor (E-mail: qinyn@ucmail.uc.edu), Department of Operations, Business Analytics and Information Systems, University of Cincinnati, Cincinnati, OH 45221.  Carey E. Priebe is Professor (E-mail: cep@jhu.edu), Department of Applied Mathematics and Statistics, Johns Hopkins University, Baltimore, MD 21210.  This work is partially supported by the XDATA program of the Defense Advanced Research Projects Agency (DARPA).  The authors thank the associate editor and two referees for insightful comments that greatly improved the article.}}
\date{\today}
\maketitle

\begin{abstract}
This article introduces a robust hypothesis testing procedure: the Lq-likelihood-ratio-type test (LqRT).  By deriving the asymptotic distribution of this test statistic, the authors demonstrate its robustness both analytically and numerically,  and they investigate the properties of both its influence function and its breakdown point.  A proposed method to select the tuning parameter $q$ offers a good efficiency/robustness trade-off, compared with the traditional likelihood ratio test (LRT) and other robust tests.  A simulation and real data analysis provides further evidence of the advantages of the proposed LqRT method.  In particular, for the special case of testing the location parameter in the presence of gross error contamination, the LqRT dominates the Wilcoxon-Mann-Whitney test and the sign test at various levels of contamination.

\end{abstract}

\section{INTRODUCTION}\label{sec:intro}
The likelihood ratio test (LRT) is one of the most frequently used statistical tools, applied in many areas of scientific research, yet its robustness is not ideal.  A robust statistical procedure performs nearly optimally when the model assumptions are valid but still performs well enough even if the assumptions are violated.  In contrast, the LRT can achieve its optimal performance only under several strict assumptions, whereas its performance degrades significantly in the presence of even mild violations of these assumptions.  In an attempt to overcome this problem, we propose a robust testing procedure, the Lq-likelihood-ratio-type test (LqRT), that leverages the newly developed concept of Lq-likelihood \citep{Ferrari2010}.  


Specifically, we consider a gross error model $h(x) = (1-\varepsilon) f(x;\theta)+\varepsilon g(x)$, where $f$ is an ``idealized'' model with the parameter $\theta$ that we need to test, $g$ is the measurement error or contamination component, and $\varepsilon$ is the contamination ratio.  $h$ represents the true data generating process, which is a deviation from $f$ when $\varepsilon>0$.  For a data set generated from $h$, the majority of the data points (i.e., $100(1-\varepsilon)\%$ portion) are drawn from $f$, whereas the rest of the data points from $g$ are considered measurement errors or outliers.  Common choices for the contamination distribution $g$ are (a) a fat tail distribution and (b) a point mass distribution.

The measurement error problem has been one of the most practical problems in Statistics.  Suppose we have some measurements $X=(X_1, X_2, ... , X_n)$ generated by a scientific experiment.  $X$ follows a distribution $f(x;\theta)$ with an interpretable parameter $\theta$, our parameter of interest.  However, we do not observe $X$, instead, we observe $X^*=(X_1^*, X_2^*, ... , X_n^*)$, where most of the $X_i^*=X_i$, but there are a few outliers due to human error, instrument malfunction, or the complexity of the underlying process.  In other words, $X^*$ is $X$ contaminated with gross errors.  Under such circumstances, using data $X^*$, we still have $\theta$ as the target parameter for our hypothesis testing or estimation \citep{Bickel2007}.  To overcome this problem, we introduce the LqRT.

For a robust hypothesis testing procedure, researchers proposed various methods with desirable properties.  \citet{Huber1965} suggested a censored likelihood ratio test (HLRT) as $T({\bf x})=\prod_{i=1}^{n}\max(c',\min(c'',p_1(x_i)/p_0(x_i)))$.  The tuning parameters $c'$ and $c''$ address the effect of outliers whose likelihood is exceedingly small and causes the ratio $p_1(x_i)/p_0(x_i)$ to approach zero or infinity.  However, hard thresholding using $c'$ and $c''$ causes maximization and minimization problems, and also induces sensitivity to the thresholds.  Accordingly, we seek to develop a smooth version of the HLRT.  Furthermore, \citet{Rousseeuw1984} proposed a ``least median of squares'' approach and corresponding testing procedures.  \citet{Heritier1994} proposed three classes of robust tests for general parametric models, namely Wald-type, score-type, and likelihood-ratio-type tests, which are the natural counterpart of M-estimators.  We will see in Section \ref{sec:LqRatioTest} that the LqRT is a member of the last class.

Meanwhile, from a robust quasi-likelihood perspective, \citet{Cantoni2001} proposed a robust deviance, which can be considered as a robust quasi-likelihood function, for hypothesis testing for generalized linear models.  \citet{Lo2009} further extended this method with saddlepoint approximations to obtain both a robust test statistic for hypothesis testing and variable selection for generalized linear models.
From a divergence point of view, \citet{Basu2013} developed a class of tests using the density power divergence (DPD) based on the work of \citet{Basu1998}.  Although they have empirically demonstrated some of strong robustness properties of the DPD-based test, their paper had no concrete theoretical results on the robustness of the proposed tests.  
In addition, \citet{Markatou1998} proposed a weighted likelihood, and \citet{Agostinelli2001} offered a test based on this weighted likelihood.  \citet{He1990} and \citet{He1991} have studied and extended the concept of a breakdown point for robustness evaluation.  
Finally, \citet{Ronchetti1997b} and \citet{Medina2015} have provided selective reviews of some basic approaches to robust inference and of recent developments in robust statistics.

Before beginning our introduction of the LqRT, we define some notations that we use throughout the text.  That is, $n$ denotes the sample size, and $d$ indicates the dimension of the observations.  Italic lower-case letters refer to scalars; for example, $x$ denotes a one-dimensional observation.  Bold-face lower-case letters refer to vectors; for example, ${\bf x}=(x_1, ..., x_n)^T$, a $n \times 1$ vector, denotes the entire sample of one-dimensional observations.  Bold-face upper-case letters refer to matrices, such as ${\bf A}$ and ${\bf B}$.  We let $p$ refer to the dimension of the parameter space, and $\theta$ indicates a $p \times 1$ parameter vector, such that $\theta_0$ denotes the true parameter.  Finally, $f(x;\theta)$ is the assumed probability density function, and $f'_{\theta}(x;\theta)$ and $f''_{\theta}(x;\theta)$ are the first and second order derivatives of the probability density function with respect to $\theta$.

In the next section, we briefly introduce the Lq-likelihood and other preliminary concepts, which lead in to the introduction of the Lq-likelihood-ratio-type test (LqRT) in Section \ref{sec:LqRatioTest}.  We demonstrate its robustness properties through an analysis of the asymptotic distribution, the influence function, and the breakdown point; we also discuss related issues such as critical values.  The numerical results are presented in Section \ref{sec:Numerical_Results}.  We then discuss the selection of $q$ in Section \ref{sec:Selectq_Lqtest} and demonstrate the superior performance of our test.  Finally, we conclude with a discussion in Section \ref{sec:Conclusion_Lqtest}, summarize the assumptions in the Appendix in Section \ref{sec:Appendix}, and relegate the proofs and additional simulation studies to online supplementary materials.

\section{PRELIMINARY CONCEPTS}\label{sec:Preliminaries}

\subsection{Lq-Likelihood and Maximum Lq-Likelihood Estimation}\label{sec:LqLikelihood_Lqtest}

A likelihood function measures the likelihood of the observed sample ${\bf x}=(x_1, ..., x_n)$ under a hypothesized model (assume a one-dimensional observation, so $d=1$).  It is defined as $\mathcal{L}({\bf x};\theta)=\prod_{i=1}^{n} f(x_i;\theta)$, where $f$ is the hypothesized model with $\theta \in \Theta \subset \mathbb{R}^p$.  Usually it is more convenient to work with the log-likelihood, $\ell({\bf x};\theta)= \sum_{i=1}^{n} \log f(x_i;\theta)$.  \citet{Ferrari2010} introduced the Lq-likelihood, defined as $\sum_{i=1}^{n} L_q( f(x_i;\theta))$.  The $L_q(\cdot)$ function with tuning parameter $q$ is defined as $L_q(u) = (u^{1-q}-1)/(1-q)$ for $q\neq1$, and $L_q(u)=\log u$ for $q=1$.  When $q \to 1$, $L_q(u) \to \log u$.  Throughout this article, we assume $0<q\leq1$.

To estimate $\theta$, we can use the maximum Lq-likelihood estimation (MLqE) defined in  \citet{Ferrari2010}: $\tilde{\theta}_{q}=\arg\max_{\theta \in \Theta} \sum_{i=1}^{n} L_q ( f(x_i;\theta))$.  To obtain $\tilde{\theta}_{q}$, we solve the Lq-likelihood equation, $0=\sum_{i=1}^{n} [f'_{\theta}(x_i;\theta)/f(x_i;\theta)] f(x_i;\theta)^{1-q}$, which is a weighted version of the likelihood equation, with the weights given by $f(x_i;\theta)^{1-q}$.  When $q<1$, data points with high likelihoods are assigned large weights.  Outliers are usually assigned small weights because of their low likelihoods, which gives the MLqE remarkable robustness.  As $q \to 1$, the MLqE becomes the maximum likelihood estimation (MLE).

The robustness added by the Lq-likelihood results because the $L_q(\cdot)$ function is bounded from below for $0<q<1$.  Note that $L_q(u)\geq-1/(1-q)$ whereas $\log(x) \to -\infty$ when $x \to 0^+$.  In this case, if an outlier such as $x_1$ gives a very small value of $f(x_1;\theta)$, then $\sum_{i=1}^{n} \log f(x_i;\theta)$ approaches $-\infty$, regardless of whether $\theta$ gives high likelihood for $x_2$, ... , $x_n$, i.e., large values of $f(x_2;\theta)$, ... , $f(x_n;\theta)$.  Furthermore, because $L_q(u)$ is bounded, it limits the effect of one particular data point on the quantity $\sum_{i=1}^{n} L_q(f(x_i;\theta))$.  Therefore, the Lq-likelihood surface is much more stable than the log-likelihood surface to any perturbation caused by a small portion of the data.

\subsection{Consistency and Bias Correction}\label{sec:BCLqLikelihood_Lqtest}
The MLqE provides a remarkably robust estimate but is consistent only for a few special cases (e.g., estimation of the location parameter of a symmetric distribution).  We consider two approaches to correct this inherent bias.

First, consider a sequence $q_n$, and let $q_n \to 1$ as $n \to \infty$, such that we have $\tilde{\theta}_{q_n} \overset{p}{\to} \theta_0$.  \citet{Ferrari2010} and \citet{Ferrari2012} offered a detailed discussion of this approach.  However, the MLqE gradually loses robustness as $q_n$ tends to 1.

Second, consider a fixed $q$ and subtract a bias correction term from the Lq-likelihood function, which is equivalent to re-centering the estimation equation.  We first have the following definition.
\begin{mydef}\label{def:bias_correction}
Define bias correction term: $C(\theta,q)=\int f(x;\theta)^{2-q}/(2-q) dx$.
\end{mydef}
With the definition above, the bias-corrected maximum Lq-likelihood estimation (BCMLqE) is defined as
\begin{align*}
\hat{\theta}_q=\arg\max_{\theta \in \Theta} \sum_{i=1}^{n} [L_q ( f(x_i;\theta))-C(\theta,q)].
\end{align*}
Section \ref{sec:Appendix} contains a simple proof for the consistency of the BCMLqE.  When $q=1$ or $\theta$ is a location parameter, $C(\theta,q)$ is a constant that is independent of $\theta$, and the BCMLqE becomes the MLqE.  We therefore adopt this bias-correction approach throughout the article, and extend it in the next section to support our proposed test statistic.  

Following this bias correction, the BCMLqE constitutes the minimum density power divergence estimator (MDPD) proposed by \citep{Basu1998}.  Therefore, maximizing the bias-corrected Lq-likelihood function is equivalent to minimizing the density power divergence between the empirical distribution and the parametric distribution.  That is, the bias-corrected Lq-likelihood function is to the density power divergence as the log-likelihood function is to the Kullback-Leibler distance.  While we are revising our paper, we became aware of another test proposed by \citet{Basu2013}.  They introduced a distance-based test using the density power divergence between the estimated model and the hypothesized model.  Their test is similar to our proposed test in spirit, but still quite different in many aspects.  First, their test is based on a single distance between the estimated parametric distribution and the hypothesized parametric distribution, whereas our test considers the difference between the maximums of likelihood functions under the null parameter space and under the union of the null and alternative parameter spaces, which is essentially a disparity difference type test and is discussed in detail in Section \ref{sec:LqRT_def}, equation \eqref{eq:LqRT_def}.  As a consequence, their test statistic only involves the data in their parameter estimate, the MDPD.  In other words, their test statistic is a function of the parameter estimate.  Therefore, the robustness of their test statistic is linked directly to the robustness of the parameter estimate.  On the other hand, our test statistic directly depends on the data itself through the likelihood functions, meaning that our test statistic makes use of information more thoroughly.  But it also complicates its robustness properties as discussed in Section \ref{sec:influence_function_breakdown_point}.  Due to this fundamental difference, the asymptotic null and alternative distributions also differ between their test and ours.  Finally, their test includes two tuning parameters, $\lambda$ and $\beta$, where $\lambda$ adjusts the distance's sensitivity to outliers and $\beta$ adjust the MDPD's robustness.  In contrast, our test only has one tuning parameter, $q$, for adjusting the Lq-likelihood function's sensitivity to outliers.

\section{Lq-LIKELIHOOD-RATIO-TYPE TEST}\label{sec:LqRatioTest}

\subsection{Test Statistic}\label{sec:LqRT_def}
We now define our proposed Lq-likelihood-ratio-type test.  Suppose we have a sample $(x_1,...,x_n)$ and an assumed parametric model $f(x;\theta)$ with the parameter $\theta \in \Theta \subset \mathbb{R}^p$.  We are interested in testing $H_0: \theta \in \Theta_0$ against $H_1: \theta \in \Theta_1$.  Here, $\Theta_0$ and $\Theta_1$ are the null and alternative parameter spaces, respectively.  We define the Lq-likelihood-ratio-type test (LqRT) as
\begin{align}\label{eq:LqRT_def}
    D_q({\bf x})&=2 \sup_{\theta \in \Theta_0 \cup \Theta_1} \Big\{\sum_{i=1}^{n} [L_q (f(x_i;\theta))- C(\theta, q)]\Big\}-2\sup_{\theta \in \Theta_0} \Big\{\sum_{i=1}^{n} [L_q( f(x_i;\theta))- C(\theta, q)]\Big\},
\end{align}
where $q$ is a tuning parameter.  We reject the null hypothesis when $D_{q}$ is large.  Moreover, the test defined by $D_{q}$ is a member of the class of likelihood-ratio-type tests as defined in \citet{Heritier1994}, equation (6), pp. 898.  It is obtained by choosing $\rho(z;\theta)=-L_{q}(f(z; \theta))+C(\theta,q)$.  Therefore, many tools provided by \citet{Heritier1994} apply directly for the LqRT.

To derive the asymptotic distribution of our test statistic, we partition the parameter as $\theta=(\alpha,\beta)$, where $\alpha \in \mathbb{R}^r$ and $\beta  \in \mathbb{R}^{p-r}$, then simplify the null and alternative hypotheses, such that they become $H_0: \alpha={\bf 0}$ and $H_1: \alpha \neq {\bf 0}$.  The test statistic is $D_{q}({\bf x})=2 \sup_{\alpha,\beta} \sum_{i=1}^{n} [L_{q_n}( f(x_i;(\alpha,\beta)))-C((\alpha,\beta),q)]-2 \sup_{\beta} \sum_{i=1}^{n} [L_{q_n}( f(x_i;({\bf 0},\beta)))-C(({\bf 0},\beta),q)]$.  The necessary definitions and assumption are as follows:
\begin{mydef}\label{def:AB_clean}
$\psi(x;\theta,q)=\frac{\partial}{\partial \theta} L_q(f(x;\theta))$,  $\psi'(x;\theta,q)=\frac{\partial^2}{\partial \theta^2} L_q(f(x;\theta))$, $c(\theta,q)=\frac{\partial}{\partial \theta}C(\theta,q)$,   $c'(\theta,q)=\frac{\partial^2}{\partial \theta^2} C(\theta,q)$, where $C(\theta,q)$ is the bias correction term introduced in Definition \ref{def:bias_correction}. 
$\tilde{\psi} (X;\theta,q)=\psi(X;\theta,q)-c(\theta,q)$, $\tilde{\psi}'(X;\theta,q)=\psi'(X;\theta,q)-c'(\theta,q)$, 
${\bf A}=\mathbb{E}[ \tilde{\psi}(X;\theta_0,q) \tilde{\psi}(X;\theta_0,q)^T ]$,
\begin{align*}
{\bf B}=-\mathbb{E}[ \tilde{\psi}'(X;\theta_0,q) ]=
\begin{pmatrix}
  {\bf B}_{\alpha\alpha} & {\bf B}_{\alpha\beta}\\
  {\bf B}_{\beta\alpha} & {\bf B}_{\beta\beta}
\end{pmatrix}, \textrm{ and }
{\bf B}^*=\begin{pmatrix}
  0 & 0 \\
  0 & {\bf B}_{\beta\beta}^{-1}
\end{pmatrix},
\end{align*}
where $\psi$, $\tilde{\psi}$, and $c$ are $p \times 1$ vectors, $\psi'$, $\tilde{\psi}'$,  $c'$ , ${\bf A}$, and ${\bf B}$ are $p \times p$ symmetric matrices.
We denote the sorted eigenvalues of an $r \times r$ matrix ${\bf M}$ by $\lambda_j({\bf M})$ for $j=1,...,r$, with $\lambda_1({\bf M}) \geq ... \geq \lambda_r({\bf M})$.  
\end{mydef}
\begin{myassump}\label{assump:interior}
$f$ satisfies the regularity conditions specified in the Appendix (Section \ref{sec:Appendix}).
\end{myassump}
\begin{mythm}\label{thm:asym_dist_BCLqRT_clean}
Under Assumption \ref{assump:interior} and a correctly specified model $f$, for a fixed $q$, the asymptotic distribution of $D_{q}({\bf x})$ under the null hypothesis is  $\sum_{j=1}^{r}\lambda_j({\bf A}[{\bf B}^{-1}-{\bf B}^*]) \chi^2_{1,j}$, where the $\chi^2_{1,j}$ are i.i.d. chi-square random variables with 1 degree of freedom, and the  $\lambda_j({\bf A}[{\bf B}^{-1}-{\bf B}^*])$ are $r$ positive eigenvalues of ${\bf A}[{\bf B}^{-1}-{\bf B}^*]$.
\end{mythm}
The proof of this result can be obtained directly as a special case of Proposition 3a, pp. 898, 899 in \citet{Heritier1994}, when $\rho(z;\theta)=-L_{q}(f(z; \theta))+C(\theta,q)$.  Also we note that when $q=1$, we have ${\bf A}={\bf B}$.  Therefore, the LqRT becomes the LRT, which follows a chi-square distribution with $r$ degrees of freedom.

\subsection{Robust Properties of LqRT}\label{sec:UnivariateCase}

To explain how the asymptotic distribution of our test changes in response to data contamination, we discuss a situation in which data are generated from a gross error model $h=(1-\varepsilon)f+\varepsilon g$, where $f$ is the assumed model and $g$ is the contamination component.  For simplicity, we assume $r=p$; that is, we restrict the case to a simple null hypothesis.  To study the robust properties, we need the following definition and assumptions:
\begin{mydef}\label{def:AB}
${\bf A}_{\varepsilon,q}=\mathbb{E}_h[ \tilde{\psi}(X;\theta^*_{\varepsilon,q},q)\tilde{\psi}(X;\theta^*_{\varepsilon,q},q)^T ]$,
and ${\bf B}_{\varepsilon,q}=-\mathbb{E}_h[ \tilde{\psi}'(X;\theta^*_{\varepsilon,q},q) ]$,
where ${\bf A}_{\varepsilon,q}$, and ${\bf B}_{\varepsilon,q}$ are $p \times p$ symmetric matrices, and $\theta^*_{\varepsilon,q}=\arg\max_{\theta} \mathbb{E}_h[L_q(f(X;\theta))-C(\theta,q)]$ and $C(\theta,q)$ is introduced in Definition \ref{def:bias_correction}.
\end{mydef}
{\it Remark:} Definition \ref{def:AB} is a generalization of Definition \ref{def:AB_clean} for gross error models, because ${\bf A}={\bf A}_{0,1}$, and ${\bf B}={\bf B}_{0,1}$.  Here, $\theta^*_{\varepsilon,q}$ represents the parameter to which the BCMLqE converges under the gross error model $h$.  In addition, $\theta^*_{0,q} = \theta_{0}$ for $0<q \leq 1$.
To study the robust properties, we assume the following conditions.  
\begin{myassump}\label{assump:contamination}
For any $\varepsilon \in (0,1)$, the gross error model $h$ is such that $\mathbb{E}_h[f''_{\theta}(X;\theta^*_{\varepsilon,1})/f(X;\theta^*_{\varepsilon,1})]$ is positive definite, where $\theta^*_{\varepsilon,1}=\arg\max_{\theta}\mathbb{E}_h[\log f(X;\theta)]$.
\end{myassump}
\begin{myassump}\label{assump:q_effect}
$\mathbb{E}_f [f'_{\theta}(X;\theta_0)f'_{\theta}(X;\theta_0)^T (f(X;\theta_0)^{-2q}-f(X;\theta_0)^{-q-1})] $ is negative definite for any $q \in (0,1)$.
\end{myassump}
\begin{myassump}\label{assump:q_monotone}
There exists a constant $q^{**} \in (0,1)$, such that $\lambda_j({\bf A}_{\varepsilon,q}{\bf B}_{\varepsilon,q}^{-1})$ are monotonic function in $q$ for any $q \in (q^{**},1)$.
\end{myassump}
{\it Remark:} A detailed discussion on these assumptions can be found in the online supplementary materials, Section \ref{sec:assumption_supp}, in which we explain the implications of these assumptions and what these assumptions mean for the exponential family in general, and for normal distributions in particular.

\begin{mythm}\label{thm:asym_dist_contamination}
Under Assumption \ref{assump:interior} and a misspecified model $h$, for a fixed $q$, the asymptotic distribution of $D_{q}({\bf x})$ under the null hypothesis is $\sum_{j=1}^{r}\lambda_j({\bf A}_{\varepsilon,q}{\bf B}_{\varepsilon,q}^{-1}) \chi^2_{1,j}$, where the $\chi^2_{1,j}$ are i.i.d. chi-square random variables with 1 degree of freedom, and the $\lambda_j({\bf A}_{\varepsilon,q}{\bf B}_{\varepsilon,q}^{-1})$ are $r$ positive eigenvalues of ${\bf A}_{\varepsilon,q}{\bf B}_{\varepsilon,q}^{-1}$.
\end{mythm}

{\it Remark:} When $\varepsilon=0$ and $q= 1$, $\lambda_j({\bf A}_{\varepsilon,q}{\bf B}^{-1}_{\varepsilon,q})=1$, and the LqRT follows a chi-square distribution with $r$ degrees of freedom.

Next, for $\lambda_j({\bf A}_{\varepsilon,q}{\bf B}_{\varepsilon,q}^{-1})$, we have the following theorem: 

\begin{mythm}\label{thm:eigenvalues_larger_1}
Under Assumptions \ref{assump:interior} and \ref{assump:contamination} and a misspecified model h, for any $\varepsilon \in (0,1)$ and for $q=1$, it holds that $\lambda_j({\bf A}_{\varepsilon,1}{\bf B}^{-1}_{\varepsilon,1}) > 1$ for $j=1,...,r$.
\end{mythm}

{\it Remark}: Theorem \ref{thm:eigenvalues_larger_1} in turn indicates that, when contamination occurs in the data and $q=1$, the divergence between ${\bf A}_{\varepsilon,1}$ and ${\bf B}_{\varepsilon,1}$ increases, so $\lambda_j({\bf A}_{\varepsilon,1}{\bf B}_{\varepsilon,1}^{-1})$ increases away from 1, causing inflation in the asymptotic distribution.  The original chi-square distribution with $r$ degrees of freedom becomes a sum of the $r$ inflated chi-square distributions with 1 degree of freedom, with each inflation captured by $\lambda_j({\bf A}_{\varepsilon,1}{\bf B}_{\varepsilon,1}^{-1})$, $j=1,...,r$.  Therefore, $D_{q}$ follows an ``inflated'' chi-square distribution under the null hypothesis.  The same phenomenon occurs for the asymptotic distribution under the alternative hypothesis, which is an ``inflated'' non-central chi-square distribution.  The overlap between the null and alternative distributions then becomes larger, and the power of the test degrades.  As the contamination becomes more serious, the overlap grows larger, and the power decreases (see Figure \ref{fig:asymptotic_dist_3dGaussian} for illustration).  To limit or control this degradation of power, we need to control for the inflation of the asymptotic distribution.  The following theorem illustrates how we can do so with $q<1$:
\begin{mythm}\label{thm:shrunk_eigenvalues}
Under Assumptions \ref{assump:interior}, \ref{assump:contamination}, \ref{assump:q_effect}, and \ref{assump:q_monotone} and a misspecified model $h$, there exists an $\tilde{\varepsilon} \in (0,1)$, such that for any $\varepsilon \in (0, \tilde{\varepsilon})$, there exists a $q^* \in (q^{**},1)$ and for any $q \in (q^*,1)$, we have
\begin{align*}
    |\lambda_j({\bf A}_{\varepsilon,q}{\bf B}_{\varepsilon,q}^{-1})-1| < |\lambda_j({\bf A}_{\varepsilon,1}{\bf B}_{\varepsilon,1}^{-1})-1| \textrm{ for } j = 1,...,r.
\end{align*}
\end{mythm}

{\it Remark:} If we were to remove Assumption \ref{assump:q_monotone} from Theorem \ref{thm:shrunk_eigenvalues}, Theorem \ref{thm:shrunk_eigenvalues} would remain true, but only for a particular $q$.  That is, we can show that there exists a $q<1$, such that $|\lambda_j({\bf A}_{\varepsilon,q}{\bf B}_{\varepsilon,q}^{-1})-1| < |\lambda_j({\bf A}_{\varepsilon,1}{\bf B}_{\varepsilon,1}^{-1})-1|$.

{\it Remark:} Theorem \ref{thm:shrunk_eigenvalues} also implies that, by setting $q<1$, we can shrink the eigenvalues $\lambda_j({\bf A}_{\varepsilon,q}{\bf B}_{\varepsilon,q}^{-1})$ back toward 1 and alleviate the inflation of distributions.  More importantly, the effect of $q<1$ on $\lambda_j({\bf A}_{\varepsilon,q}{\bf B}_{\varepsilon,q}^{-1})$ can offset the inflation effect of $\varepsilon>0$ on $\lambda_j({\bf A}_{\varepsilon,q}{\bf B}_{\varepsilon,q}^{-1})$.  With this approach, we can avoid the increasing overlap between the null and alternative distributions and protect the power of the test better.

In summary, the divergence between ${\bf A}_{\varepsilon,q}$ and ${\bf B}_{\varepsilon,q}$ due to contamination is more serious for $q=1$ than $q<1$.  Even though ${\bf A}_{0,1}={\bf B}_{0,1}$ at zero contamination, the loss of power at $\varepsilon>0$, due to the divergence between ${\bf A}_{\varepsilon,1}$ and ${\bf B}_{\varepsilon,1}$, is not avoidable.  On the other hand, setting $q<1$ would lead to the loss of the exact equality at zero contamination, i.e., ${\bf A}_{0,q}\neq {\bf B}_{0,q}$, but the divergence between ${\bf A}_{\varepsilon,q}$ and ${\bf B}_{\varepsilon,q}$ is much less with substantial contamination, hence it preserves the test's power.  In other words, by setting $q<1$, we trade the exact equality of ${\bf A}_{0,1}={\bf B}_{0,1}$ for much less divergence between ${\bf A}_{\varepsilon,q}$ and ${\bf B}_{\varepsilon,q}$ at $\varepsilon>0$.

The theory offers two key insights.  First, our test statistic makes the approximation ${\bf A}_{\varepsilon,q} \approx {\bf B}_{\varepsilon,q}$ more robust to model misspecification. By setting $q<1$, many of the statistical inferences originally based on ${\bf A}_{0,1}={\bf B}_{0,1}$ can still remain valid, even if the model is misspecified.  Second, we gain a tool for identifying model misspecification.  Setting $q<1$ effectively eliminates the influence of outliers, and setting $q>1$ can magnify the effect of outliers.  Therefore, the difference between ${\bf A}_{\varepsilon,q}$ and ${\bf B}_{\varepsilon,q}$ for $q>1$ will be more sensitive to model misspecification.  When $q=1$, ${\bf A}_{\varepsilon,1}$ is essentially Fisher's information matrix.  Many model misspecification tests are based on ${\bf A}_{\varepsilon,1}={\bf B}_{\varepsilon,1}$ \citep[e.g.,][]{White1982}.  The preceding results also provide a possible approach to model misspecification detection.

\subsection{Simulation Study}\label{sec:sim_mean_normal_Asymptotic_Multivariate}
In this section, we present simple simulations to support the findings in the previous section.

\begin{figure}[h]
\centering
    \includegraphics[scale=0.4]{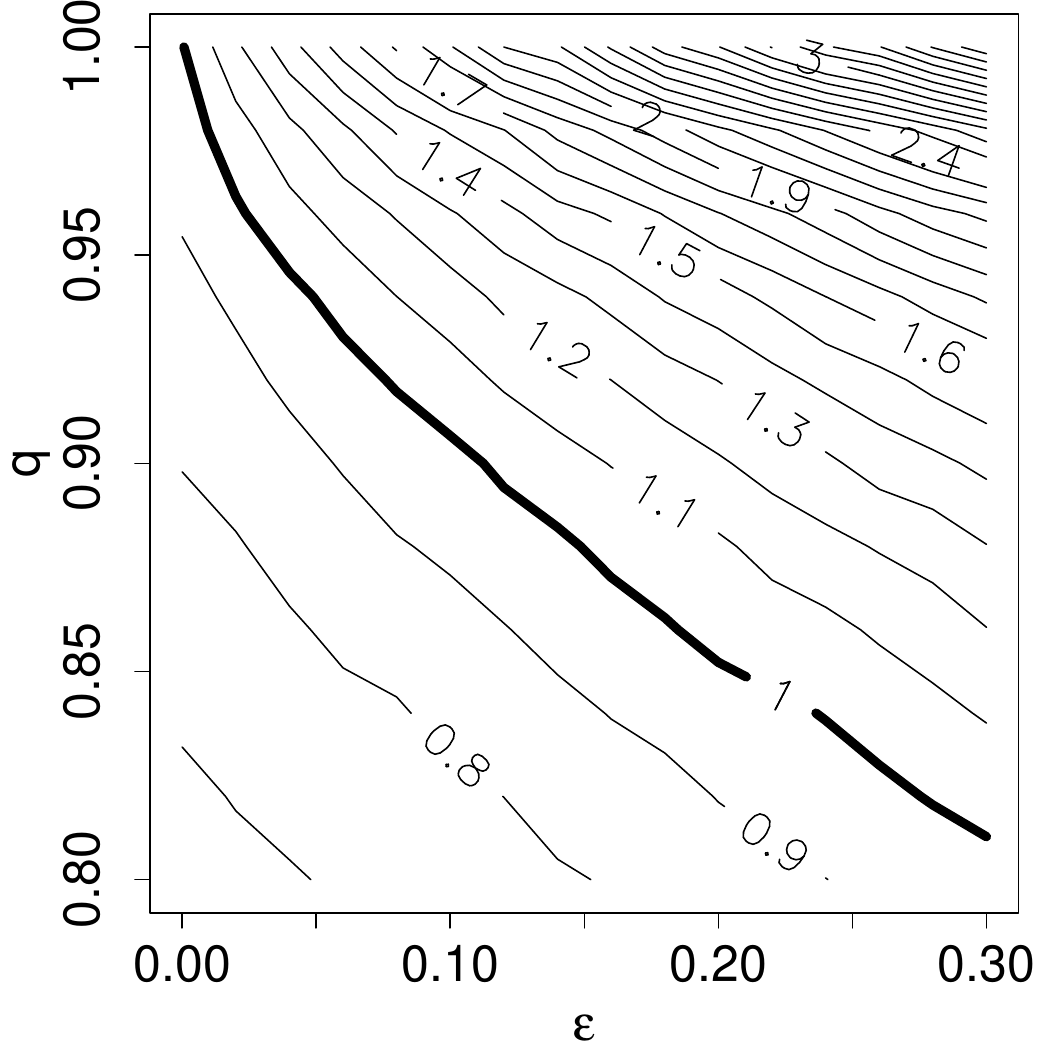}
    \caption{Contour plot of $\lambda({\bf A}_{\varepsilon,q}{\bf B}_{\varepsilon,q}^{-1})$ as a function of $\varepsilon$ and $q$ where the assumed model is a standard normal distribution.}
    \label{fig:ratio_at_q_vs_eps}
\end{figure}

First, we plot $\lambda({\bf A}_{\varepsilon,q}{\bf B}_{\varepsilon,q}^{-1})$ as a function of the contamination ratio $\varepsilon$ and the tuning parameter $q$ for $p=1$ (i.e., only one eigenvalue) in Figure \ref{fig:ratio_at_q_vs_eps}.  In this case, $f$ is a normal distribution. The data generating process is $h(x)=(1-\varepsilon) \varphi(x;0,1) + \varepsilon \varphi(x;0,10)$.   In Figure \ref{fig:ratio_at_q_vs_eps}, we highlight the contour level of 1 in bold.  As the figure shows, $\lambda({\bf A}_{\varepsilon,q}{\bf B}_{\varepsilon,q}^{-1})$ increases as $\varepsilon$ increases when $q=1$.  However, we can always find a value of $\lambda({\bf A}_{\varepsilon,q}{\bf B}_{\varepsilon,q}^{-1})$ that is closer to 1 by decreasing $q$.

\begin{figure}
    \includegraphics[scale=0.54]{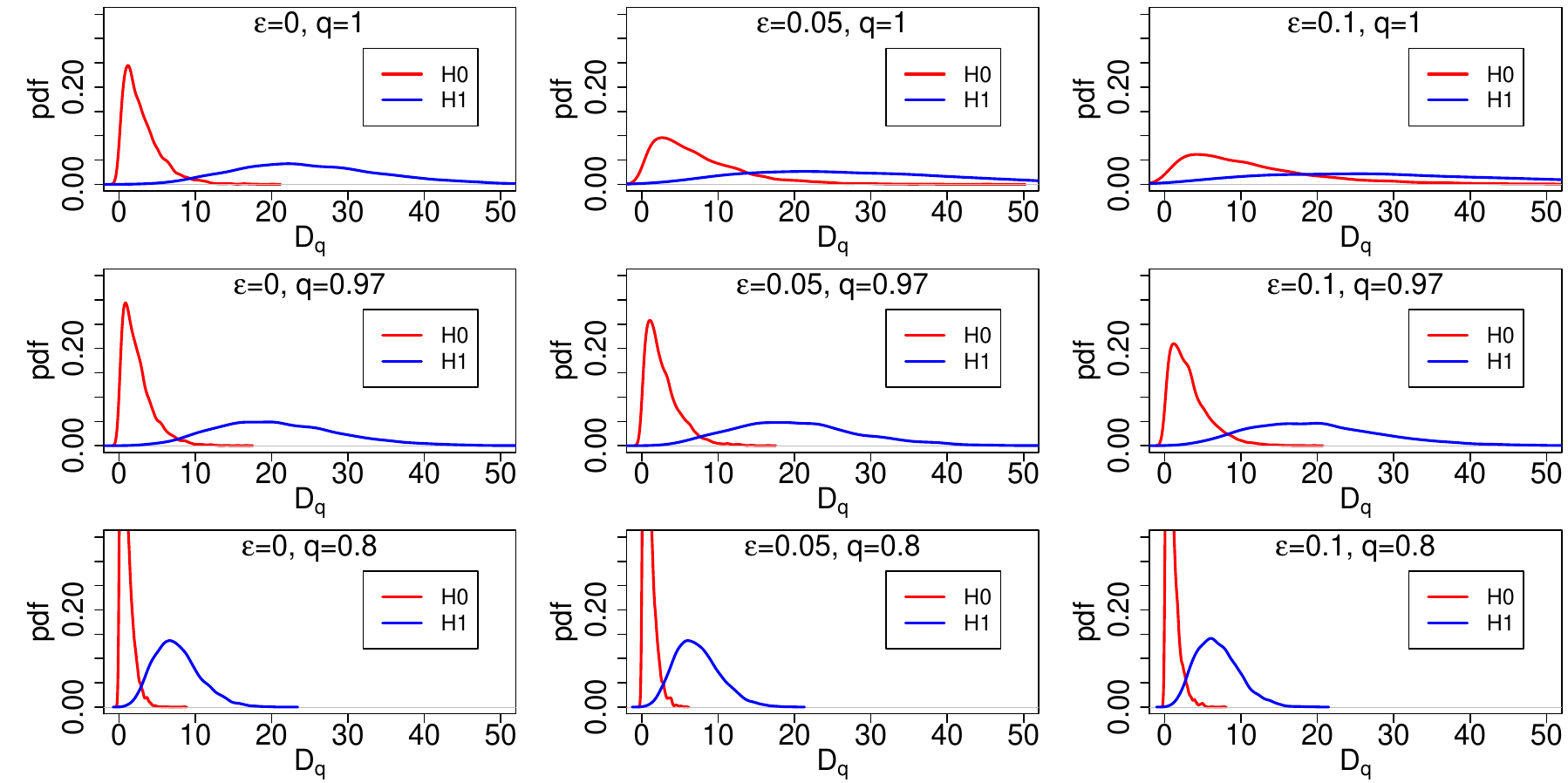}
    \caption{Comparison of asymptotic null and alternative distributions of the test statistic for testing the mean of the three-dimensional normal distribution under difference levels of symmetric heavy-tail contamination and different $q$s.}
    \label{fig:asymptotic_dist_3dGaussian}
\end{figure}

Second, we simulate the asymptotic null and alternative distributions under $\varepsilon=0, 0.05$, and $0.1$ and $q=1, 0.97$, and $0.8$ for $H_0: \mu_f=0$ and $H_1: \mu_f \neq 0$.  For this test, $f$ is a three-dimensional normal distribution with known variance.  The data generating process is $h=(1-\varepsilon) f + \varepsilon g$, where $g$ is another multivariate normal distribution with $\mu_g=\mu_f$ and $\Sigma_g=30\Sigma_f$.  We simulate the null and alternative distributions of $D_q$ using $\mu_f=[0, 0, 0]^T$ and $\mu_f=[0.15, 0.15, 0.15]^T$ and present the results in Figure \ref{fig:asymptotic_dist_3dGaussian}.

In this figure, when $q=1$ and $\varepsilon$ increases from 0 to 0.1, the null and alternative distributions become flatter and overlap more, which results in power degradation.  When $q=0.97$, instead of having the inflated chi-square distribution, the null and alternative distributions are less affected by the contamination, because $\lambda_j({\bf A}_{\varepsilon,q}{\bf B}_{\varepsilon,q}^{-1})$ gets pulled back toward 1 when we set $q<1$.  When $q=0.8$, the distributions are much less affected.  It is worth noting that, in the case of $q=0.8$ and $\varepsilon=0$ (lower left panel), the null and alternative distributions overlap more than they do in the case of $q=1$ and $\varepsilon=0$ (upper left panel), which means that by setting $q<1$, we lose some of the test's power at zero contamination.  As Figure \ref{fig:asymptotic_dist_3dGaussian} illustrates, we gain robustness from using the Lq-likelihood and make a trade-off for robustness by giving up a little power at zero contamination.

\subsection{Critical Values}\label{sec:Critical_Value}
Since we know the asymptotic null distribution of LqRT from Theorem \ref{thm:asym_dist_contamination}, we can calculate the $1-\alpha$ quantile of the null distribution to obtain the critical value for a level $\alpha$ test.  When $\varepsilon=0$, we use the following algorithm to calculate the critical value:

Step 1: Calculate ${\bf A}_{0,q}$ and ${\bf B}_{0,q}$ under $H_0$.

Step 2: Calculate $\lambda_j({\bf A}_{0,q} {\bf B}^{-1}_{0,q})$.

Step 3: Obtain the $1-\alpha$ quantile of the distribution of $\sum_{j=1}^{r} \lambda_j({\bf A}_{0,q} {\bf B}^{-1}_{0,q}) \chi^2_{1,j}$ using either simulation or the tools provided by \citet{Rao1981} and \citet{Modarres1992}.  Denote it as $\text{CV}_{\alpha}$.  \citet{Rao1981} proposed using a linear transformation of $\chi^2_{r}(1-\alpha)$, the $1-\alpha$ quantile of the chi-square distribution with $r$ degrees of freedom, to approximate the $1-\alpha$ quantile of the distribution of $\sum_{j=1}^{r} \lambda_j({\bf A}_{0,q} {\bf B}^{-1}_{0,q}) \chi^2_{1,j}$.  For example, we could use $[\sum_{j=1}^{r}\lambda_j({\bf A}_{0,q} {\bf B}^{-1}_{0,q})/r] \chi^2_{r}(1-\alpha)$.  We found it works well in practice, especially when $\lambda_j$ is close to 1.  If we suppose $\lambda_1=0.9$, $\lambda_2=0.85$, $\lambda_3=0.8$, then the critical value obtained from the simulation of 100,000 iterations is 6.639, and the critical value obtained from $[\sum_{j=1}^{r}\lambda_j({\bf A}_{0,q} {\bf B}^{-1}_{0,q})/r] \chi^2_{r}(1-\alpha)$ is 6.643.  Note that the simulation could calculate the true critical value accurately, but since $\alpha$ is usually small, it takes a relatively long time.

We use $\text{CV}_{\alpha}$ as the critical value.  Even though $\text{CV}_{\alpha}$ is intended for the case of $\varepsilon=0$, it also works well for the case of $\varepsilon>0$, as long as $q$ is not very close to 1.  The reason is that the null and alternative distributions of LqRT do not change drastically as $\varepsilon$ increases away from 0 when $q<1$ (Figure \ref{fig:asymptotic_dist_3dGaussian}).  So the $1-\alpha$ quantile of the null distribution at $\varepsilon=0$ is very close to the $1-\alpha$ quantile of the null distribution at $\varepsilon>0$.  In Section \ref{sec:sim_covariance}, we also confirm that this approach works well for mild contamination.  Unfortunately, it does not work very well with heavy contamination, because the null distribution varies too much due to that contamination.

Can we then obtain the genuine critical value for the case of $\varepsilon>0$?  The answer is yes...but not easily.  The difficulty is due to the fact that the null distribution depends on $\varepsilon$ and $g$ (Theorem \ref{thm:asym_dist_contamination}), which are never known in practice.  For example, $g$ can be an arbitrary distribution, which makes it very difficult to estimate.  Without a good estimate of $g$, it is also hard to estimate $\varepsilon$.  Therefore, it can be challenging to obtain the true $1-\alpha$ quantile of the null distribution for $\varepsilon>0$.  We next present some special cases that allow us to estimate the critical values, though in general, we would use $\text{CV}_{\alpha}$.

\subsubsection{Location Parameter}\label{sec:Critical_Value_Location}
To test a location parameter, we can use the bootstrap method to estimate the critical value from the sample. Suppose $H_0:\theta=\theta_0$ and $H_1:\theta\neq\theta_0$, where $\theta$ is the location parameter.  We propose the following algorithm:

Step 1: Given a sample ${\bf x}=(x_1, ..., x_n)$, we estimate the mean using a robust procedure, such as the BCMLqE $\hat{\theta}_{q}$.

Step 2: Shift the entire sample by $\theta_0-\hat{\theta}_{q}$ to obtain ${\bf x'}=(x_1-\hat{\theta}_{q}+\theta_0, ..., x_n-\hat{\theta}_{q}+\theta_0)$.

Step 3: Use ${\bf x'}$ to get bootstrap samples ${\bf x}'_b$ for $b=1, ... , B$.

Step 4: Calculate $D_q({\bf x}'_b)$ for each bootstrap sample and denote each one as $D^b_{q}$.

Step 5: Calculate the $1-\alpha$ quantile of $D^b_{q}$.  Denote it as $\widehat{\text{CV}}_{\alpha,q}$.

As a result, $\widehat{\text{CV}}_{\alpha,q}$ is our estimate for the critical value.  The rationale behind our method is as follows.  We first transform the sample ${\bf x}$ to have a mean of $\theta_0$.  With this new sample ${\bf x'}$, we use the bootstrap to mimic the null distribution.  Since there are usually outliers in the sample, we choose a robust estimation for the mean, namely, the BCMLqE.  We demonstrate this approach in Section \ref{sec:sim_mean_normal}.

\subsubsection{Linear Regression}\label{sec:Critical_Value_Regression}
Assume a linear regression setting, $y_i = {\bf x}_i^T \beta + \eta_i$, where $y_i \in \mathbb{R}$, ${\bf x}_i, \beta \in \mathbb{R}^p$, and $\eta_i$ is the error term.  To test $H_0:\beta=\beta_0$ and $H_1:\beta \neq \beta_0$, we propose a similar algorithm.

Step 1: Given a sample $\{y_i, {\bf x}_i\}_{i=1,...,n}$, we first obtain a robust estimate $\hat{\beta}$.  $\hat{\beta}$ can be obtained through BCMLqE.  In addition, other choices that provide robust initial estimates of the coefficient can also be used, for example, the ridge regression and the least absolute deviations (LAD) regression.

Step 2: Calculate the residual $\hat{\eta}_i=y_i-{\bf x}^T_i \hat{\beta}$ for $i=1,...,n$.

Step 3: Use $\{\hat{\eta}_i\}_{i=1,...,n}$ to get bootstrap samples $\{\hat{\eta}^b_i\}_{i=1,...,n}$ for $b=1, ... , B$.

Step 4: Obtain $y^b_i={\bf x}^T_i \beta_0+\hat{\eta}^b_i$ for $i=1,...,n$ using $\beta_0$.

Step 5: Calculate $D_q(\{y^b_i, {\bf x}_i\}_{i=1,...,n})$ for each bootstrap sample and denote each one as $D^b_{q}$.

Step 6: Calculate the $1-\alpha$ quantile of $D^b_{q}$.  Denote it as $\widehat{\text{CV}}_{\alpha}$.

Thus, $\widehat{\text{CV}}_{\alpha}$ is our estimate for the critical value.  The rationale again is that we want to transform the sample so that it preserves the contamination as much as possible (Step 3) and satisfies $H_0$ (Step 4).  We apply this approach in Section \ref{sec:sim_linear_regression}.

\subsection{Influence Function and Breakdown Point}\label{sec:influence_function_breakdown_point}
We now turn to the analysis of the influence function and the breakdown point.  In this section, we use $F$ and $G$ to denote the distribution functions of $f$ and $g$, and we consider all the test statistics as statistical functionals with domain $\mathcal{F}$, the set of all proper distributions.  

The influence function \citep{Hampel1986} pertains to the infinitesimal behavior of the real-valued functional.  It measures the effect of an infinitesimal contamination at the point $x$ on the estimator, so it can be considered as a proxy for the asymptotic bias caused by the contamination at $x$.  \citet{Ronchetti1979,Ronchetti1982a,Ronchetti1982} have further extended the influence function to hypothesis testing by defining a level influence function (LIF) and a power influence function (PIF).  They show how the asymptotic level and power are influenced by a small amount of contamination at a particular point.  For our test statistic $D_q$, since it is not Fisher-consistent (i.e., $D_q(F_{\theta}) \neq \theta$), we can modify the test statistic by defining $U(G)=\xi^{-1}(D_q(G))$ where $\xi(\theta)=D_q(F_\theta)$, so that $U(G)$ is Fisher-consistent \citep{Huber2009}.  However, the properties of the influence function are relatively more difficult to determine for the likelihood-ratio-type test, compared with the Wald or score test statistics, because the LqRT test statistic involves the data through both the parameter estimate and the Lq-likelihood function \citep{Basu2013}.  For ease of presentation, we mainly study the influence function of another closely related test statistic, $T=\hat{\theta}_q$, which is the BCMLqE of $\theta$.  That is, we focus on $T$, which is Fisher-consistent and used directly in $D_q$.  The LIF of $T$ is
\begin{align*}
\textrm{\textrm{LIF}}(x;T,F)=\frac{\phi(\Phi^{-1}(1-\alpha_0))\textrm{IF}(x;T,F)}{\sqrt{\int \textrm{IF}(x;T,F)^2 dF(x)}},
\end{align*}
where $\alpha_0$ is the nominal level of the test.  Therefore, $\textrm{LIF}(x;T,F)$ is proportional to $\textrm{IF}(x;T,F)$, the influence function.  We know that $\textrm{IF}(x;T,F)$ is proportional to $\tilde{\psi}$.  For most of the parametric models that satisfy the regularity conditions, $\tilde{\psi}$ is bounded.  Therefore, both $\textrm{IF}$ and $\textrm{LIF}$ are bounded.  These properties add stability to our test, in terms of both level and power.

The boundedness of the influence function can be decided on a case-by-case basis.  We illustrate a general method for deciding the boundedness of the influence function for our test.  For the BCMLqE, the influence function is proportional to $\tilde{\psi}$,
\begin{align*}
\tilde{\psi}(x;\theta,q)=\frac{f'_{\theta}(x;\theta)}{f(x;\theta)}f(x;\theta)^{1-q}-\int f'_{\theta}(x;\theta)f(x;\theta)^{1-q} dx,
\end{align*}
where the second term is independent of $x$.  Therefore, the boundedness depends on only the first term.  When $f$ belongs to the exponential family, we have 
\begin{align*}
\tilde{\psi}(x;\theta,q)=(T(x)-A'_{\theta}(\theta))f(x;\theta)^{1-q}-\int f'_{\theta}(x;\theta)f(x;\theta)^{1-q} dx.
\end{align*}
We know that $f(x;\theta) \to 0$ as $x \to \infty$, and that $(T(x)-A'_{\theta}(\theta))f(x;\theta)^{1-q}$ is a continuous function in $x$.  Therefore, as long as $(T(x)-A'_{\theta}(\theta))f(x;\theta)^{1-q} \to 0$ as $x \to \infty$, we have the boundedness of the influence function, such as a normal distribution, exponential distribution, and so on.

We now move to the analysis of the breakdown point.  Intuitively, the breakdown point of a statistical functional is the fraction of data that can be given arbitrary values without making the statistical functional arbitrarily bad.  For a test, the concept of the breakdown point can be naturally extended to the level breakdown point (LBP) and power breakdown point (PBP).  Let $t_{\text{max}}=\sup_{F \in \mathcal{F}}D_q(F)$ and $t_{\text{min}}=\inf_{F \in \mathcal{F}}D_q(F)$, and define level breakdown function $\varepsilon_0$ and power breakdown function $\varepsilon_1$ as
\begin{align*}
\varepsilon_0(F_\theta,D_q)=\inf\{\varepsilon: \sup_{G \in \mathcal{F}} D_q((1-\varepsilon)F_{\theta}+\varepsilon G) = t_{\text{max}} \},\\
\varepsilon_1(F_\theta,D_q)=\inf\{\varepsilon: \inf_{G \in \mathcal{F}} D_q((1-\varepsilon)F_{\theta}+\varepsilon G) = t_{\text{min}} \}.
\end{align*}
Intuitively, $\varepsilon_0(F_\theta,D_q)$ represents the smallest amount of contamination necessary to drive the p-value of $D_q$ to 0, which leads to the level breakdown.  Similarly, $\varepsilon_1(F_\theta,D_q)$ represents the smallest amount of contamination necessary to drive the p-value of $D_q$ to 1, which leads to the power breakdown.  If $\theta \in \Theta_1$ and $t_{\text{min}}=0$, then $\varepsilon_1(F_\theta,D_q)$ is the smallest fraction of contamination that can make the LqRT inconsistent.  Finally, we define the level breakdown point (LBP) and power breakdown point (PBP) as
\begin{align*}
\text{LBP}(D_q)=\sup_{\theta \in \Theta_0} \varepsilon_0(F_\theta,D_q),\\
\text{PBP}(D_q)=\sup_{\theta \in \Theta_1} \varepsilon_1(F_\theta,D_q).
\end{align*}

\begin{mythm}\label{thm:breakdown_point_LqLR}
The level breakdown point and power breakdown point of LqRT are the same as the breakdown point of the BCMLqE, $\hat{\theta}_q$.
\end{mythm}
{\it Remark:} The exponential family satisfies all the regularity conditions of the BCMLqE.  Furthermore, the influence function of the BCMLqE on the parameter in this case is bounded (i.e., $\tilde{\psi}$ is bounded), therefore, the BCMLqE has a breakdown point of 0.5.  Consequently, the LqRT has both the LBP and PBP of 0.5 for the exponential family.

\section{NUMERICAL RESULTS}\label{sec:Numerical_Results}
We present numerical studies illustrating the performance of our proposed method.  Sections \ref{sec:sim_mean_normal},  \ref{sec:sim_linear_regression} and \ref{sec:sim_covariance} focus on simulation results.  Section \ref{sec:Real_Data} focuses on real data analysis.  Additional simulation studies can also be found in the online supplementary materials.

\subsection{Mean of Normal Distribution}\label{sec:sim_mean_normal}

\begin{figure}
    \includegraphics[scale=0.46]{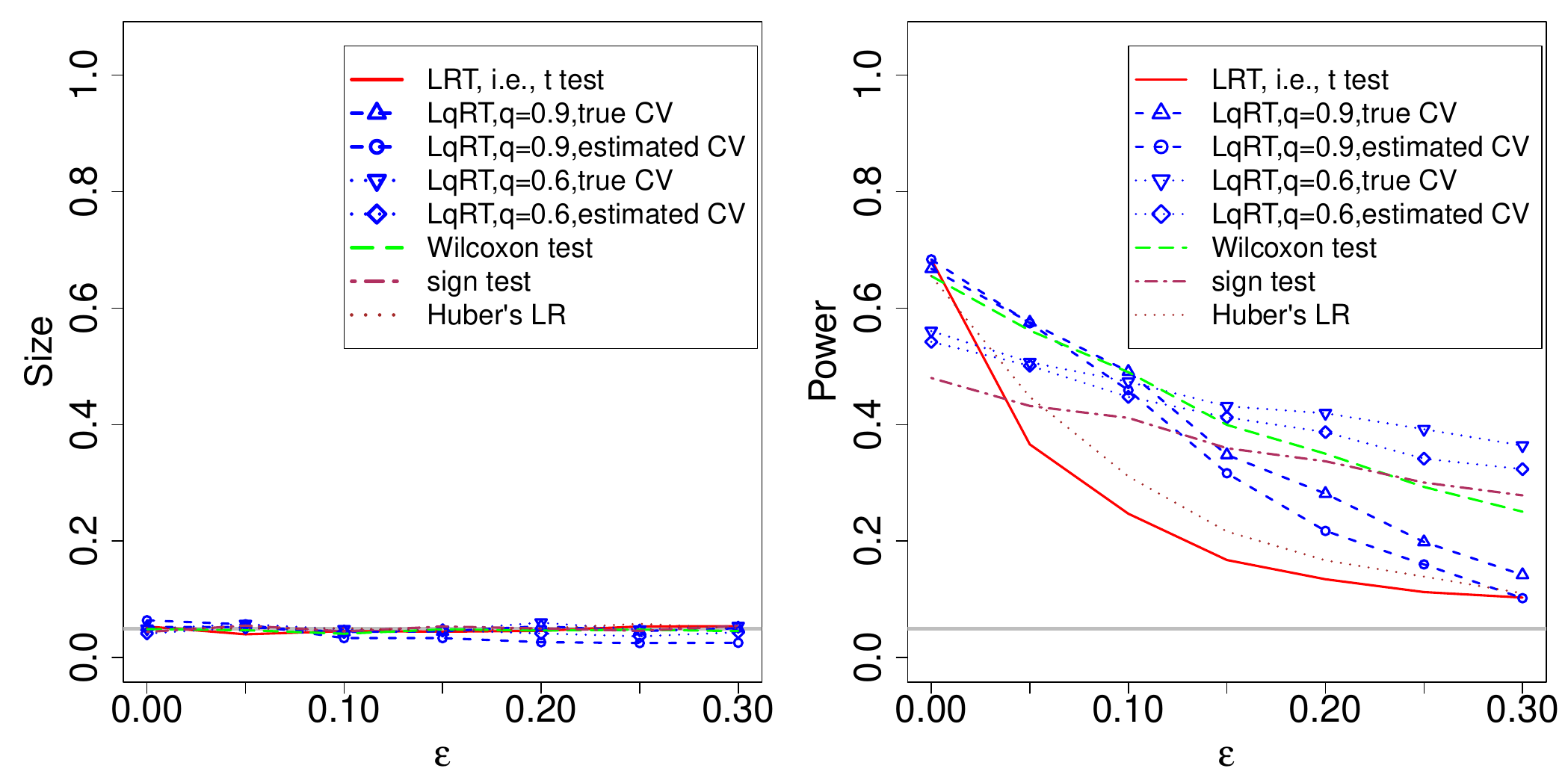}
    \caption{Comparison of powers and sizes for the LqRT ($q=0.9, 0.6$), LRT, Wilcoxon test, sign test, and HLRT at different levels of heavy-tail contamination when testing for the mean of the normal distribution ($H_0: \theta = 0$, $H_1: \theta \neq 0$).  The powers are calculated using the data generating process with mean $\theta=0.34$.}
    \label{fig:Power_comparison_unknown_variance}
\end{figure}

We assume $f$ to be a normal distribution with unknown mean $\theta$ and variance $\sigma^2$, and we test $H_0: \theta = 0$ against $H_1: \theta \neq 0$.  We simulate data with a sample size of $n=50$ from $h(x;\theta,\varepsilon)=(1-\varepsilon)\varphi(x;\theta,1)+\varepsilon \varphi(x;\theta,50)$.  Then we apply the LqRT with $q=0.9$ and $0.6$, the likelihood ratio test (LRT) which is equivalent to t test, the Wilcoxon test, the sign test, and the Huber's censored likelihood ratio test (HLRT) with $c'=0.1$ and $c''=10$.  At different levels of $\varepsilon$, we use $h(x;\theta=0,\varepsilon)$ and $h(x;\theta=0.34,\varepsilon)$ to generate the data and calculate the size and power.  We use the approach introduced in Section \ref{sec:Critical_Value_Location} to generate critical values.  The results are in Figure \ref{fig:Power_comparison_unknown_variance}.  

In the left panel of Figure \ref{fig:Power_comparison_unknown_variance}, the sizes of all tests are successfully controlled at 0.05, which indicates that the estimated critical values work well.  In the right panel of Figure \ref{fig:Power_comparison_unknown_variance}, at zero contamination, the LRT achieves the highest power;  the LqRT with $q=0.9$, the Wilcoxon test, and the HLRT also offer high powers.  As contamination becomes more serious, the LRT degrades much faster than any of the other tests.  The LqRT with $q=0.9$ and $0.6$ both degrade at much slower rates.  The Wilcoxon test also degrades slowly.  Among all tests, the LqRT with $q=0.6$ and the sign test offer the slowest degradation rates.  By setting $q$ to $0.9$ and $0.6$, we can beat the Wilcoxon test at mild ($\varepsilon<0.05$) and heavy ($\varepsilon>0.15$) contamination, respectively.  Meanwhile, the LqRT with $q=0.6$ uniformly dominates the sign test at all levels of contamination.  Finally, we see that we only slightly overestimate the critical values as the powers obtained from the estimated critical values are slightly below these of the true critical values.

In addition, we perform simulation studies using different alternative models (i.e., $h(x;\theta=0.2,\varepsilon)$, $h(x;\theta=0.5,\varepsilon)$, and $h(x;\theta=0.8,\varepsilon)$) and different contamination (i.e., point mass contamination) and obtain similar results.  Please see the online supplementary materials, Section \ref{sec:add_sim_supp}, for detail.

\subsection{Linear Regression}\label{sec:sim_linear_regression}
\begin{figure}  \centering
    \includegraphics[scale=0.45]{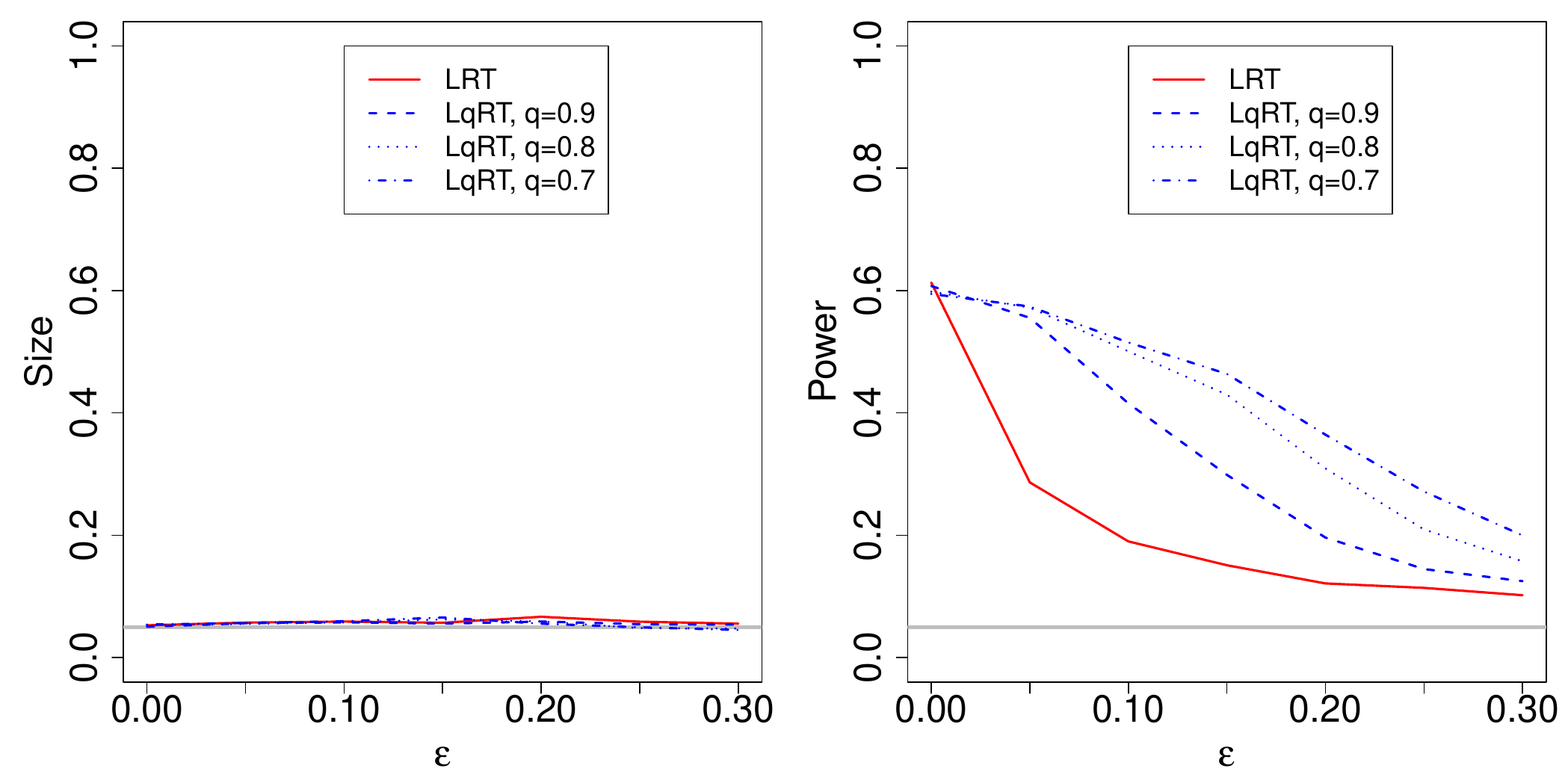}
    \caption{Comparison of powers and sizes for the LqRT ($q=0.9, 0.8, 0.7$) and the LRT under different levels of heavy-tail contamination when testing for the coefficients of the linear regression model ($H_0: \beta_1=\beta_2$, $H_1: \beta_1 \neq \beta_2$).}
    \label{fig:Power_comparison_reg_fixed_q}
\end{figure}

We also test the proposed method in a linear regression setting.  Suppose the assumed model is $y_i= \beta_1 x_{i1} + \beta_2 x_{i2}+ \eta_i$, where $\eta_i \sim \varphi(0,\sigma^2)$, and we want to test $H_0: \beta_1=\beta_2$ against $H_1: \beta_1 \neq \beta_2$.  With a sample size of $n=100$, we simulate data using $x_{i1} \sim \text{Uniform}(0,0.5)$, $x_{i2} \sim \text{Uniform}(0,0.5)$, $\eta_i \sim (1-\varepsilon) \varphi(0,0.2)+\varepsilon \varphi(0,10)$, and $y_i= 1 x_{i1} + 1 x_{i2}+ \eta_i$, and calculate the size of the test.  In addition, we simulate data according to $y_i= 0.5 x_{i1} + 1.5 x_{i2}+ \eta_i$ to calculate the power.  We compare the LqRT ($q=0.9,0.8,0.7$) with the LRT.  We use the approach introduced in Section \ref{sec:Critical_Value_Regression} for critical values.  The results are given in Figure \ref{fig:Power_comparison_reg_fixed_q}.

In Figure \ref{fig:Power_comparison_reg_fixed_q}, we see that all tests successfully control the sizes at 0.05, which means estimated critical values work well.  As for the power, at $\varepsilon=0$, the power of LqRT is slightly lower than the power of LRT.  In contrast, as $\varepsilon$ increases, the LqRT degrades much more slowly than the LRT and has much higher power than the LRT when $\varepsilon>0$.  

\subsection{Covariance}\label{sec:sim_covariance}
\begin{figure}  \centering
    \includegraphics[scale=0.45]{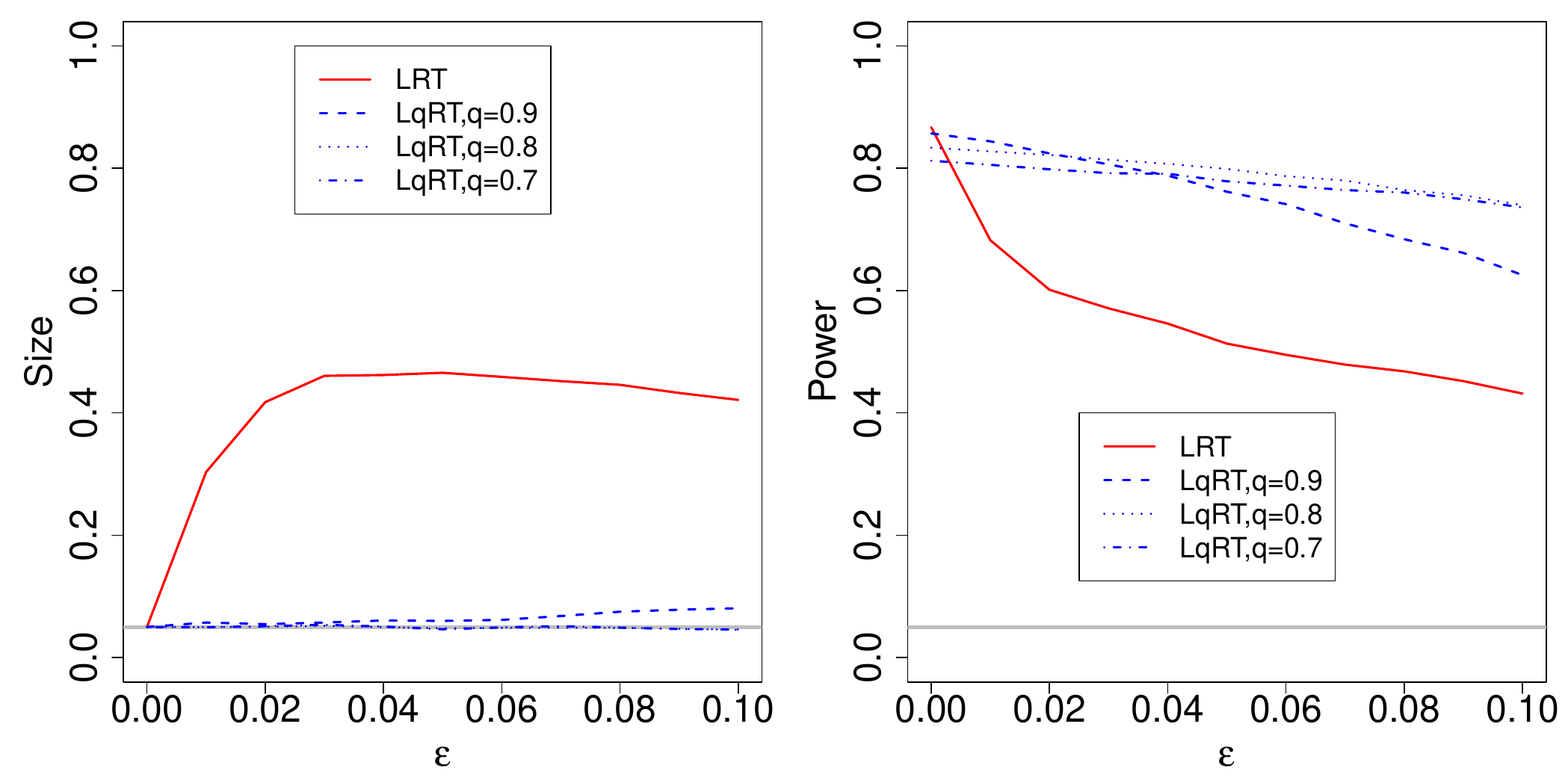}
    \caption{Comparison of powers and sizes for the LqRT ($q=0.9, 0.8, 0.7$) and the LRT under different levels of heavy-tail contaminations when testing for the covariance matrix of the multivariate normal distribution ($H_0: \Sigma_{1,2}=0$, $H_1: \Sigma_{1,2}\neq 0$).}
    \label{fig:Power_comparison_covariance}
\end{figure}

We apply the proposed method to test for the covariance matrix too.  Suppose the assumed model is ${\bf x} \sim \text{MN}({\bf 0},{\bf \Sigma})$, where ${\bf x} \in \mathbb{R}^{2}$, ${\bf \Sigma}=[\Sigma_{i,j}] \in \mathbb{R}^{2 \times 2}$, and $\Sigma_{i,j}=\text{Cov}(X_i,X_j)$.  We test $H_0: \Sigma_{1,2}=0$ against $H_1: \Sigma_{1,2}\neq 0$.  With a sample size of $n=100$, we simulate the data using ${\bf x} \sim (1-\varepsilon)\text{MN}({\bf 0},{\bf I})+\varepsilon \text{MN}({\bf 0},30{\bf I})$ and calculate the size of the test.  We further change the data generating process to ${\bf x} \sim (1-\varepsilon)\text{MN}({\bf 0},{\bf \Sigma})+\varepsilon \text{MN}({\bf 0},{\bf \Sigma}^*)$ with $\Sigma_{1,1}=\Sigma_{2,2}=1$, $\Sigma^*_{1,1}=\Sigma^*_{2,2}=30$, and $\Sigma_{1,2}=\Sigma_{2,1}=\Sigma^*_{1,2}=\Sigma^*_{2,1}=0.3$, and simulate data to calculate the power of the test.  We use the general approach proposed in Section \ref{sec:Critical_Value} to obtain the critical values.  The results are presented in Figure \ref{fig:Power_comparison_covariance}.

As Figure \ref{fig:Power_comparison_covariance} reveals, as $\varepsilon$ increases, the size of LRT increases drastically above 0.05, and the sizes of LqRT tests increase only slightly.  The general approach for the critical value thus appears to work.  However, if we were to extend $\varepsilon$ to 0.3, the size of LqRT also would increase significantly.  That is, the general approach only works for mild contamination.  With regard to the power, we again see that it degrades relatively slowly for the LqRT compared with the LRT, which is consistent with the right panel of Figure \ref{fig:Power_comparison_reg_fixed_q}.

\subsection{Real Data}\label{sec:Real_Data}
\begin{figure}
\centering
\begin{minipage}{.5\textwidth}
    \includegraphics[scale=0.4]{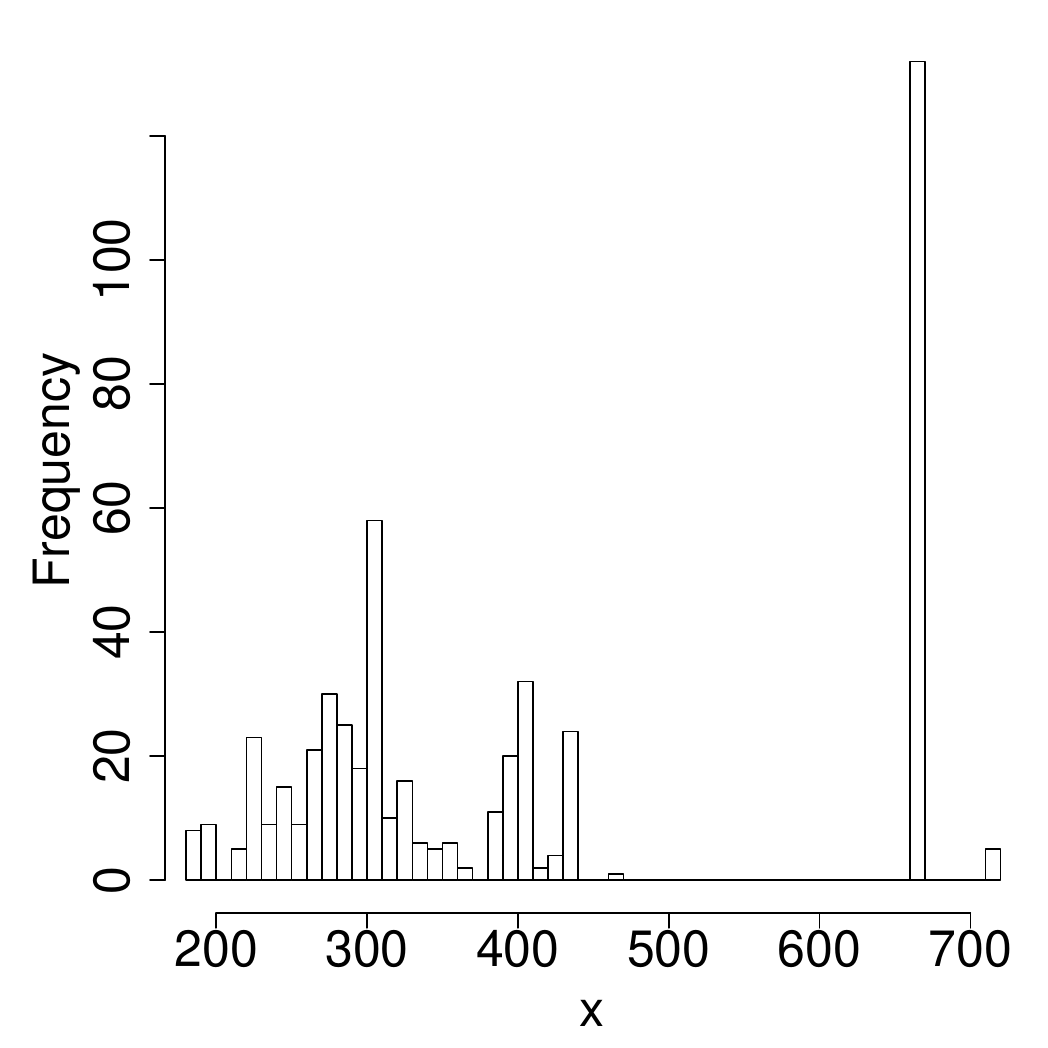}
    \caption{Histogram of property-tax rates.}
    \label{fig:hist_real_data}
\end{minipage}%
\begin{minipage}{.5\textwidth}
    \includegraphics[scale=0.4]{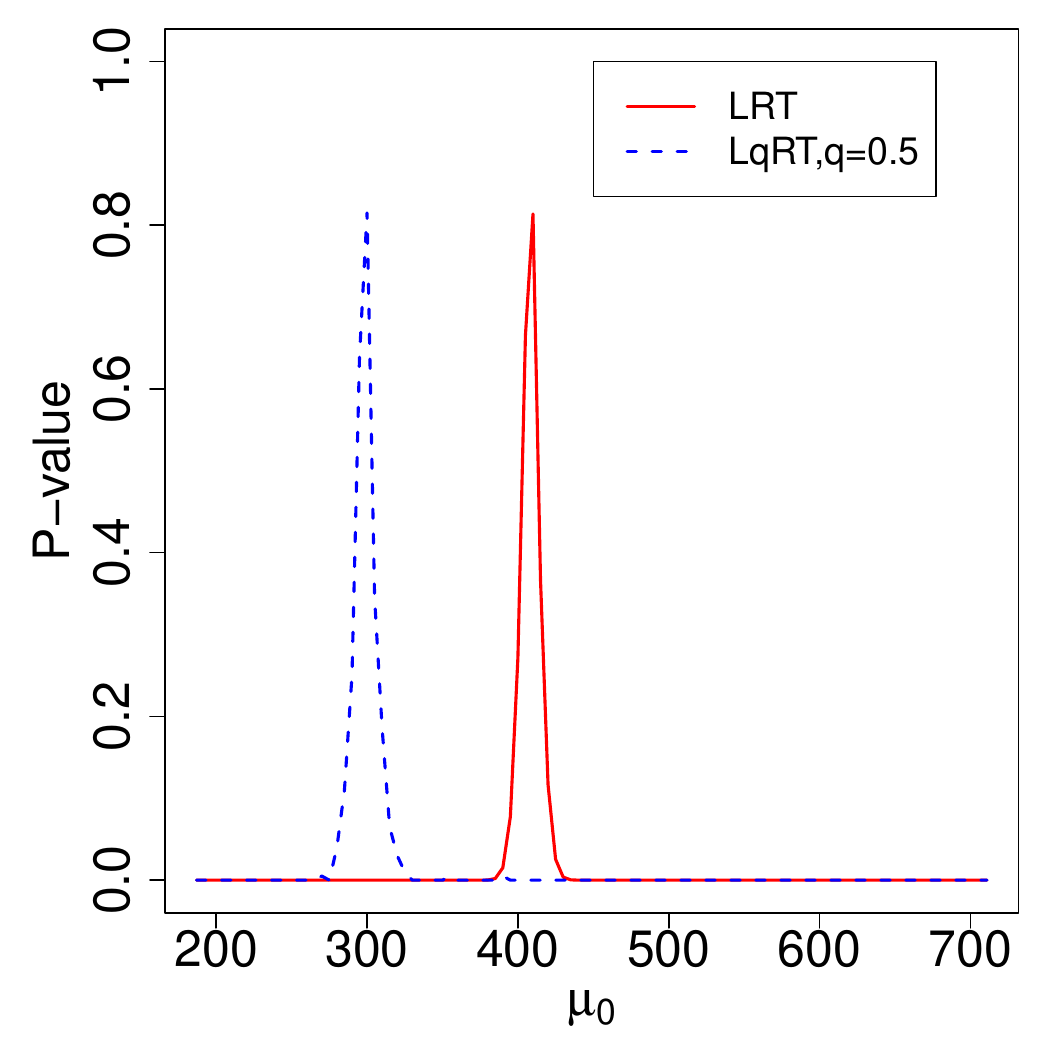}
    \caption{p-values as a function of $\mu_0$ for the LqRT and the LRT.}
    \label{fig:pvalue_real_data}
\end{minipage}
\end{figure}

We apply our test to the Boston housing data set (\url{https://archive.ics.uci.edu/ml/datasets/Housing}).  The sample size is $n=506$.  The variable ``full-value property-tax rate per \$10,000'' serves as our variable of interest.  By plotting the histogram in Figure \ref{fig:hist_real_data}, we clearly see outliers above 600.  The mean of the entire data set, including these outliers, is 408.2, whereas the mean of the data without outliers is 311.9.  Here, 311.9 offers a more reasonable estimate of the true center of data.  With this data set, we perform hypothesis testing with $H_0: \mu=\mu_0$ and $H_1: \mu \neq \mu_0$.  By varying $\mu_0$ from 200 to 700, we can plot the corresponding p-values for the LqRT ($q=0.5$) and the LRT in Figure \ref{fig:pvalue_real_data}.  For the LqRT, the p-value jumps above 0.05 at around 300, which means we reject the null hypothesis as long as $\mu_0$ is distant from the true center of the data.  The p-value of LRT instead jumps above 0.05 at around 400, and the LRT rejects the null hypothesis when $\mu_0$ is at 311.9.  It is true that in this example one could first remove outliers and then proceed with standard (non-robust) inference methods, but our preference, for the reasons noted in \citet{Huber2009} pp. 4-5, is for robust methods.

\section{SELECTION OF $q$}\label{sec:Selectq_Lqtest}

\begin{figure}  \centering
    \includegraphics[scale=0.4]{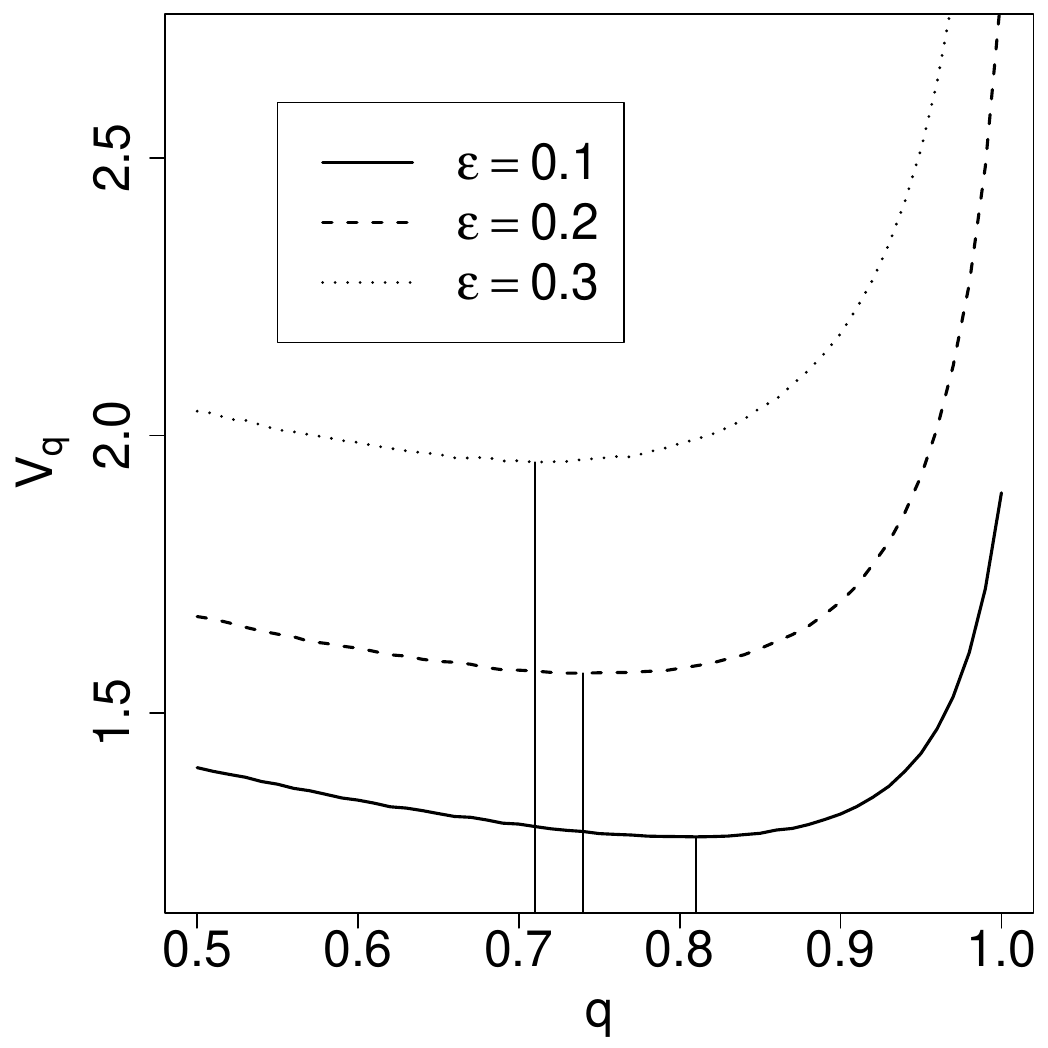}
    \caption{$V_q(\theta_0)$ as a function of $q$ at different levels of $\varepsilon$.}
    \label{fig:select_q_Lqtest}
\end{figure}

Thus far, we have assumed the tuning parameter $q$ to be known, but we never know the optimal $q$ in practice.  Therefore, in this section, we propose a method for adaptively selecting $q$.  The optimal $q$ is defined $q_{\textrm{opt}}=\arg\min_{q} 
\text{trace}( V_q(\theta_0))$, where $V_q(\theta_0)$ is the asymptotic variance of $\hat{\theta}_q$.  In Figure \ref{fig:select_q_Lqtest}, we plot the relationship between $V_q(\theta_0)$ and $q$ at different levels of contamination using the same setup as in Section \ref{sec:sim_mean_normal}.  The optimal $q$ is generally between 0.6 and 0.9. The more serious the contamination, the lower the optimal $q$.  In practice, using the empirical variance $\hat{V}_q(\hat{\theta}_q)$, we propose the data-adaptive estimation for the tuning parameter
\begin{align*}
    \hat{q}=\arg\min_{q} \text{trace}\bigg( \bigg[\frac{1}{n}\sum_{i=1}^n\tilde{\psi}'(x_i;\hat{\theta}_q,q)\bigg]^{-1}
    \bigg[\frac{1}{n}\sum_{i=1}^n \tilde{\psi}(x_i;\hat{\theta}_q,q)\tilde{\psi}(x_i;\hat{\theta}_q,q)^T\bigg]
    \bigg[\frac{1}{n}\sum_{i=1}^n\tilde{\psi}'(x_i;\hat{\theta}_q,q)\bigg]^{-1} \bigg).
\end{align*}

\begin{figure}
    \includegraphics[scale=0.46]{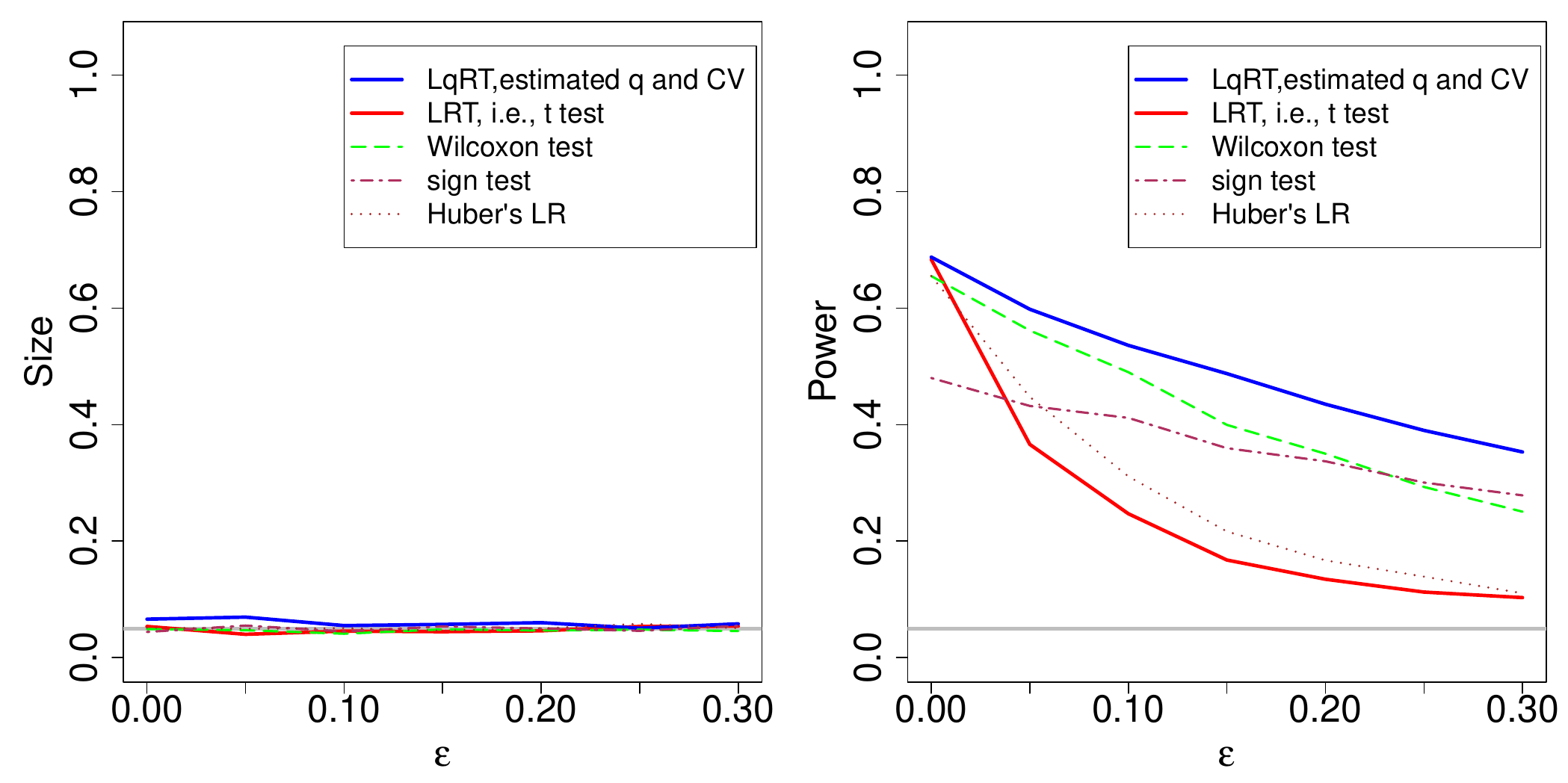}
    \caption{Comparison of the powers and sizes for the LqRT with estimated $q$ and estimated critical value, LRT, Wilcoxon test, sign test, and HLRT under different levels of heavy-tail contamination when testing for the mean of the normal distribution ($H_0: \theta = 0$, $H_1: \theta \neq 0$).  The powers are calculated using the data generating process with mean $\theta=0.34$.}
    \label{fig:comparison_est_q}
\end{figure}

Accordingly, we conduct some simulation studies of the LqRT using the estimated $q$.  We first adopt the same setup from Section \ref{sec:sim_mean_normal}.  By setting $\theta$ to 0 and 0.34, we compare the sizes and powers of our test, the LRT, the Wilcoxon test, the sign test, and the HLRT.  The results are presented in Figure \ref{fig:comparison_est_q}.

In Figure \ref{fig:comparison_est_q}, at $\varepsilon=0$, the LRT offers the highest power, though our LqRT provides nearly the same power.  As $\varepsilon$ increases away 0, the LRT's power quickly drops below all other tests.  Our test shows remarkable robustness and it degrades slower than the Wilcoxon test.  The power of our test also dominates both the Wilcoxon and sign tests uniformly at all levels of contamination.  Therefore, not only does the LqRT preserve efficiency almost perfectly at $\varepsilon=0$, but it also obtains robustness comparable to that of the nonparametric tests.  The reason our test beats the nonparametric tests is that we can adaptively control the amount of information used, by selecting $q$, whereas the Wilcoxon and sign tests always use the rank information.

\begin{figure}  \centering
    \includegraphics[scale=0.5]{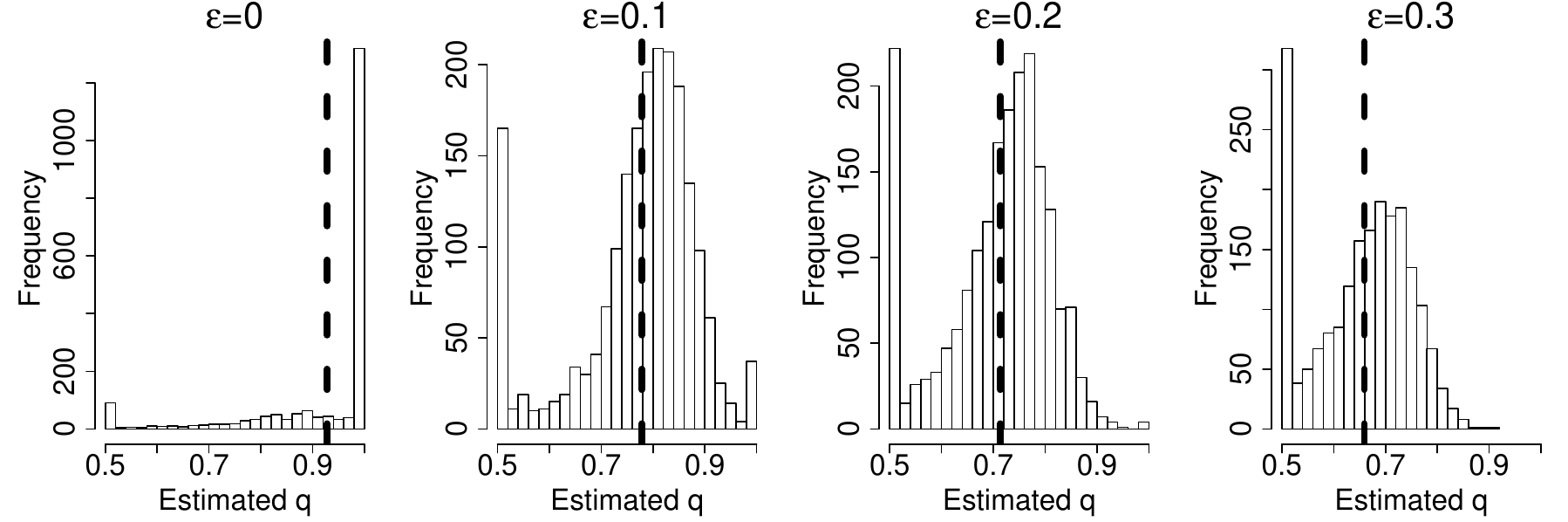}
    \caption{Histograms of the estimated $q$ of the LqRT at different levels of heavy-tail contamination when testing for the mean of the normal distribution.  These estimated $q$s are obtained when calculating the powers (i.e., the right panel of Figure \ref{fig:comparison_est_q}).  The mean estimated $q$ is indicated by a vertical dashed line.}
    \label{fig:hist_est_q}
\end{figure}

Note that Figure \ref{fig:comparison_est_q} shows the average power over 2000 Monte Carlo iterations.  Since each iteration has a different $q$, we plot these estimated $q$s in the histograms in Figure \ref{fig:hist_est_q}.  As the contamination grows more serious, the estimated $q$ tends to become smaller.  In our experiment, we limit the smallest $q$ to 0.5 (which corresponds to the minimum Hellinger distance estimation \citep{Beran1977}) because we have not understood the case of $q<0.5$ very well.  Comparing Figures \ref{fig:comparison_est_q} and \ref{fig:hist_est_q} with Figure \ref{fig:Power_comparison_unknown_variance}, we can see that the LqRT with estimated $q$ can combine the advantages of the LqRTs with fixed $q$s.  When the contamination ratio is low, the estimated $q$s tend to be large, so the LqRT with estimated $q$ has roughly the same performance as the LqRT with $q=0.9$.  When the contamination ratio is 0.3, the mean estimated $q$ is near 0.6; therefore, the LqRT with estimated $q$ achieves a performance comparable to the LqRT with $q=0.6$.

\begin{figure}  \centering
    \includegraphics[scale=0.45]{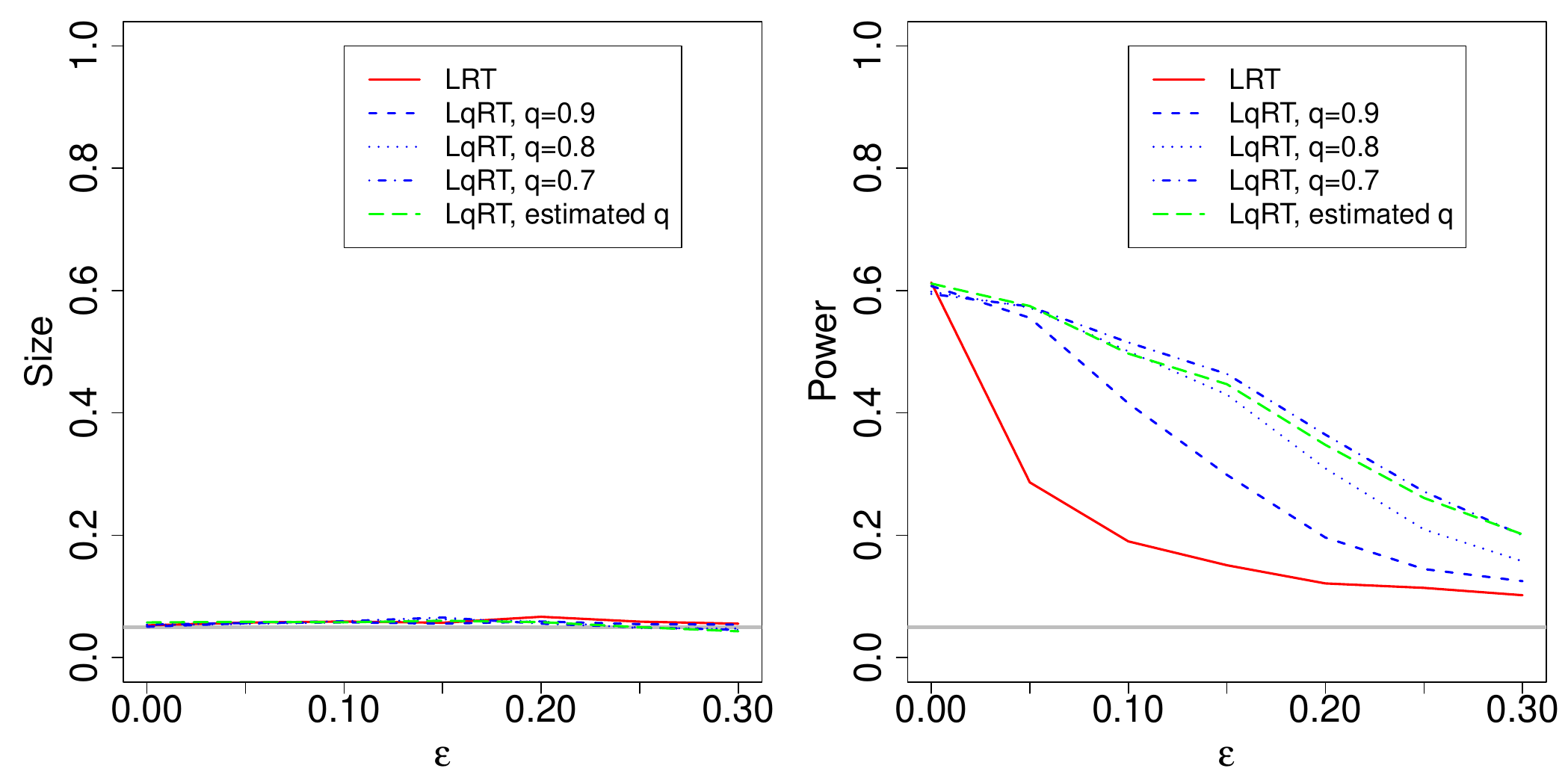}
    \caption{Comparison of powers and sizes for the LqRT with estimated $q$ and estimated critical value and the LRT under different levels of heavy-tail contamination when testing for the coefficients of the linear regression model ($H_0: \beta_1=\beta_2$, $H_1: \beta_1 \neq \beta_2$).}
    \label{fig:Power_comparison_reg_selected_q}
\end{figure}

We next repeat the simulation from Section \ref{sec:sim_linear_regression} for linear regression using LqRT with estimated $q$ and present the results in Figure \ref{fig:Power_comparison_reg_selected_q}.  As we can see, while the size is successfully controlled, the power of LqRT with estimated $q$ degrades the most slowly, and maintains relatively high compared with other LqRT tests with fixed $q$.

Through these simulation studies, we demonstrated the improved performance using estimated $q$.  In our experience, the method for estimating $q$ works well when the model is relatively simple.  When the model becomes more complex, there are more parameters to estimate.  The diagonal elements of $V_q(\theta_0)$ usually involve different scales.  Simply minimizing the trace will lead to an unsatisfactory estimated $q$.  Furthermore, when the model becomes more complex, $\hat{V}_q(\hat{\theta}_q)$ becomes unreliable, which also leads to an unsatisfactory estimated $q$.

Additional simulation studies on estimated $q$ can also be found in the online supplementary materials, Section \ref{sec:add_sim_supp}.

\section{CONCLUSION}\label{sec:Conclusion_Lqtest}

In this article, we have proposed a robust testing procedure, the Lq-likelihood ratio test (LqRT), and demonstrated its advantages over the LRT, the Wilcoxon test, the sign test, and the Huber's censored likelihood ratio test (HLRT).  We prove our test's robustness advantages by deriving the asymptotic distribution, the influence function, and the breakdown point.  We further accompany our analytical study with numerical comparisons.

In a sense, our proposed test offers a bridge, connecting the LRT with the nonparametric tests such as the Wilcoxon and sign tests.  By changing the tuning parameter $q$, we can control the information used in the hypothesis testing.  The LRT uses the full information about all data points and assigns all data points equal weights.  The Wilcoxon test takes only the rank information, and therefore achieved extreme robustness but at the cost of substantial loss of information.  Our proposed test instead assigns each data point a weight as a function of its likelihood and $q$.  Therefore, the data points that are consistent with the model earn higher weights, whereas the data points that are inconsistent with the model are partially ignored.

To the extent that the robustness of the Wilcoxon test (minimum asymptotic relative efficiency (ARE) of the Wilcoxon test versus the t test is 0.864) suggests that the Wilcoxon test should be the default test of choice (so rather than ``use Wilcoxon if there is evidence of non-normality,'' the default position should be ``use Wilcoxon unless there is good reason to believe the normality assumption''), these new results in this article suggest that the LqRT has the potential to become the new default go-to test for practitioners.

Even with the remarkable robustness of our LqRT, many directions remain for further research.  For example, researchers should seek better estimation procedures for the critical value and $q$.  Our estimate of the critical value performs decently, but there is clearly a gap in the powers obtained from the true versus the estimated critical values (see Figure \ref{fig:Power_comparison_unknown_variance}).  Filling this gap will be a challenging task.  We also need a more robust procedure for estimating $q$.  Finally, the divergence of ${\bf A}$ and ${\bf B}$, as described in Section \ref{sec:UnivariateCase},  indicates a potential approach to model misspecification detection.

\section{APPENDIX AND SUPPLEMENTARY MATERIALS}\label{sec:Appendix}

In this article, we have made the following assumptions.

{\bf Assumption \ref{assump:interior}}
For any $q \in (0,1]$, $f$ satisfies the following regularity conditions:
\begin{enumerate}[nolistsep]
\item
$\theta_0$ is an interior point in $\Theta_0$.
\item
$\sup_{\theta \in \Theta_0}\|\frac{1}{n}\sum_{i=1}^{n}\tilde{\psi}(X_i;\theta,q)-\mathbb{E}\tilde{\psi}(X;\theta,q)\|\overset{p}{\to} 0$ as $n \to \infty$, where $\|\cdot\|$ represents the $\ell_2$-norm.
\item
$\max_{1 \leq k \leq p} \mathbb{E}_{\theta_0}|\tilde{\psi}_k(X_i;\theta_0,q)|^3$, $k=1,...,p$ is upper bounded by a constant, where $\tilde{\psi}_k$ is the $k$-th element of $\tilde{\psi}$.
\item
The smallest eigenvalue of ${\bf A}$ is bounded away from zero.
\item
Let $b_{jk}$ be the $j$-th row, $k$-th column element in ${\bf B}$, then $b_{jk}^2$ for $j,k=1,...,p$ are upper bounded by a constant.
\item
The second order partial derivatives of $\tilde{\psi}(x;\theta,q)$ are dominated by an integrable functions with respect to the true distribution of $X$ for all $\theta$ in a neighborhood of $\theta_0$.
\end{enumerate}

{\bf Assumption \ref{assump:contamination}}
For any $\varepsilon \in (0,1)$, the gross error model $h$ is such that $\mathbb{E}_h[f''_{\theta}(X;\theta^*_{\varepsilon,1})/f(X;\theta^*_{\varepsilon,1})]$ is positive definite, where $\theta^*_{\varepsilon,1}=\arg\max_{\theta}\mathbb{E}_h[\log f(X;\theta)]$.

{\bf Assumption \ref{assump:q_effect}}
$\mathbb{E}_f [f'_{\theta}(X;\theta_0)f'_{\theta}(X;\theta_0)^T (f(X;\theta_0)^{-2q}-f(X;\theta_0)^{-q-1})] $ is negative definite for any $q \in (0,1)$.

{\bf Assumption \ref{assump:q_monotone}}
There exists a constant $q^{**} \in (0,1)$, such that $\lambda_j({\bf A}_{\varepsilon,q}{\bf B}_{\varepsilon,q}^{-1})$ are monotonic function in $q$ for any $q \in (q^{**},1)$.

Please see the online supplementary materials for detailed discussion on these assumptions, additional simulation studies and proofs.

\bibliographystyle{apalike}
\bibliography{YichenBib}
\end{document}